# OConsent - Open Consent Protocol for Privacy and Consent Management with Blockchain

Subhadip Mitra
WILP, Birla Institute of Technology
and Science, Pilani
contact@subhadipmitra.com

***Abstract*— In the current connected world - Websites, Mobile Apps, IoT Devices collect a large volume of users' personally identifiable activity data. These collected data is used for varied purposes of analytics, marketing, personalization of services, etc. Data is assimilated through site cookies, tracking device IDs, embedded JavaScript, Pixels, etc. to name a few. Many of these tracking and usage of collected data happens behind the scenes and is not apparent to an average user. Consequently, many Countries and Regions have formulated legislations (e.g., GDPR, EU) - that allow users to be able to control their personal data, be informed and consent to its processing in a comprehensible and user-friendly manner.**

**This paper proposes a protocol and a platform based on Blockchain Technology that enables the transparent processing of personal data throughout its lifecycle from capture, lineage to redaction. The solution intends to help service multiple stakeholders from individual end-users to Data Controllers and Privacy Officers. It intends to offer a holistic and unambiguous view of how and when the data points are captured, accessed, and processed. The framework also envisages how different access control policies might be created and enforced through a public blockchain including real time alerts for privacy data breach.**

***Keywords*— *Privacy, Blockchain, Distributed Ledger, Bitcoin, Ethereum, GDPR, Privacy, Security, PII, Cryptography***

## I. Introduction

Analysis of the Users' activity and behaviour on the websites and mobile apps provide unique insights to help businesses improve their products, service offerings and general user experience. Users' privacy and trust are key for any successful business - and thus user's consent must be sought before their data is used to maintain the said sustained trust and transparency. Given the volume of web traffic, geographies, prevalent sovereign privacy laws and multiple ways that the data points are used (e.g. Analytics, Recommendations, A/B Testing and personalization, Conversion tracking, Marketing Automation, Remarketing and User Feedback) - it is important to design a unified, open and extensible framework for Privacy and Consent Management. The framework must be able to capture consent, track lineage and enforce redaction (when consent is withdrawn).

Blockchains (and Distributed Ledger Technologies) by their very design provide trust and immutability of data . These two key features provide the building blocks of such Technology enabled Privacy Framework.

### A. *Current landscape of Data and Privacy Legislations around the world*

The relevance for such privacy controls has become important as there has been growing legislations that enforce Data Privacy around the world.

List of Data Privacy Legislations by country:

a. General Data Protection Regulation - GDPR[i] in EU,
b. Personal Data Protection Act 2012[ii] (PDPA) in Singapore,
c. Personal Data (Privacy) Ordinance[iii] in Hong Kong,
d. Federal Trade Commission - FTC & Children's Online Privacy Protection Act of 1998 (COPPA)[iv] - that enforce Data Privacy.
e. The Personal Data Protection Bill[v], 2019, India[1]
f. PIPEDA - the Personal Information Protection and Electronic Documents Act, 2000[vi], Canada
g. General Principles of Civil Law and the Tort Liability Law[vii], 1987[2], China
h. Cybersecurity Law of People's Republic of China, 2017[viii], China[ix]
i. Russian Federal Law on Personal Data, No. 152-FZ[x] and Federal Law, No. 242-FZ[xi]
j. Personal Data Protection Law, Law 19,628/1999[xii] and further amendments, Chile
k. Federal Privacy Act 1988[xiii] (Commonwealth) (Privacy Act) and Australian Privacy Principles (APPs)[xiv], Australia
l. Privacy Laws by Australian states and territories -
    - Information Privacy Act 2014 (Australian Capital Territory)[xv],
    - Information Act 2002 (Northern Territory)[xvi],
    - Privacy and Personal Information Protection Act 1998 (New South Wales)[xvii],
    - Information Privacy Act 2009 (Queensland)[xviii],
    - Personal Information Protection Act 2004 (Tasmania)[xix], and

[1] The Personal Data Protection Bill, 2019 was introduced in Lok Sabha by the Minister of Electronics and Information Technology, Mr. Ravi Shankar Prasad, on December 11, 2019. As of this writing, the Standing Committee Report has sought extension up to second week of the Winter Session 2020.

[2] Adopted Apr 12, 1986, at the 4th Session of the 6th National People's Congress, to take effect on Jan 1, 1987

- Privacy and Data Protection Act 2014 (Victoria)[xx]

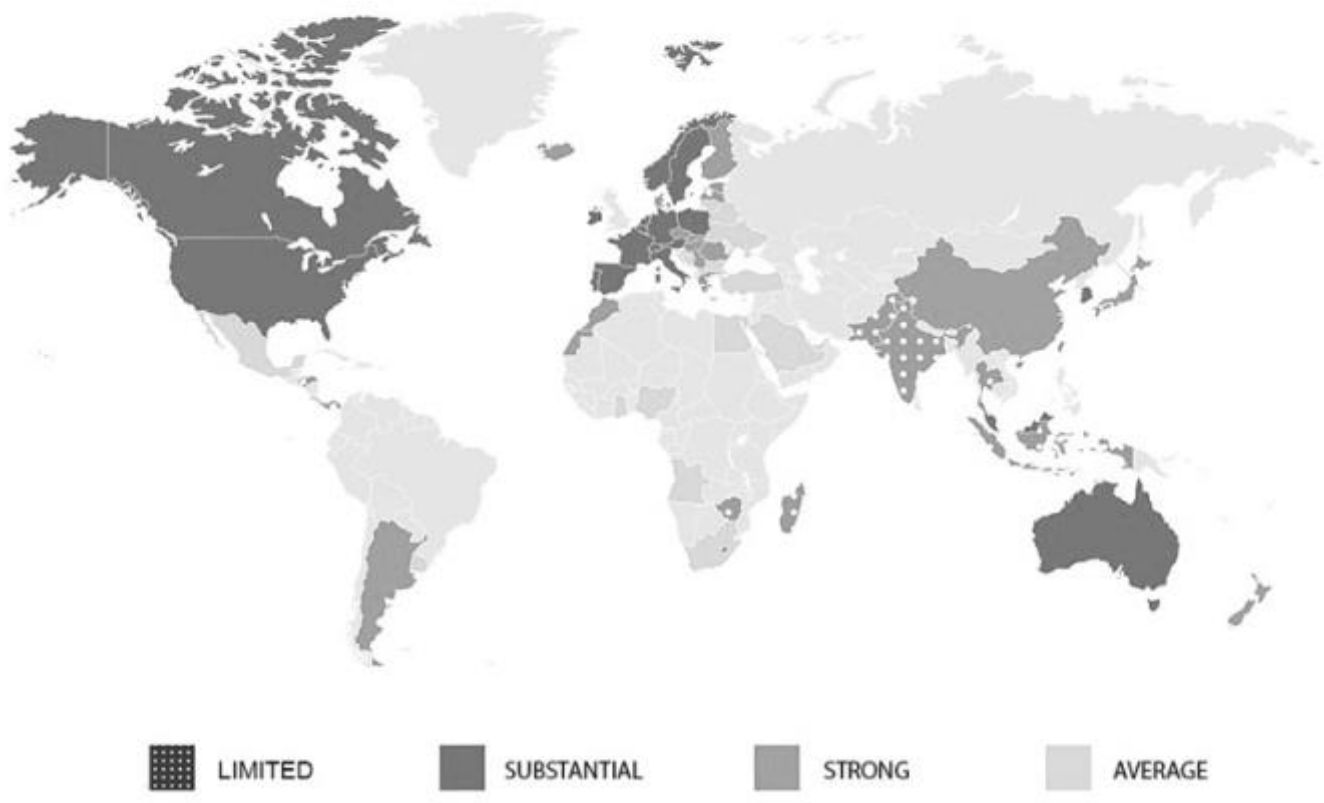

*Figure 1. Data Privacy legislative framework around the world and their maturity*

*B. Consent Definitions*

Following are the key consent definitions as per GDPR and DPA. GDPR is considered the foremost [xxi] and all-encompassing[xxii] regulation for Data Privacy and Consent Management that is modelled by other legislations across different geographies[xxiii]. Throughout this paper, discussions are aligned to GDPR regulations.

**1995 DPA Definition[xxiv]**

"... any freely given specific and informed indication of his wishes by which the data subject signifies his agreement to personal data relating to him being processed[xxv]"

**The GDPR definition**

"... any freely given, specific, informed and **unambiguous** indication of the data subject's wishes by which he or she, **by a statement or by a clear affirmative action**, signifies agreement to the processing of personal data relating to him or her[xxvi]"

*C. Consent Lifecycle*

**Stage 1: Collection -** Consent is first collected from the Data Subject (DS).

**Stage 2: Storage -** Collected consent is then securely stored.

**Stage 3: Process -** The stored consent then is processed based on the context that it was obtained for by Data Controller (DC) and a Data Processor (DP).

**Stage 4: Modification -** Consent may be modified to accommodate a change in scope.

**Stage 5: Revocation -** Consent may be revoked by DS owing to expiry or agreement breach.

**Stage 6: Archive -** Consent data may be archived for regulatory and audit needs.

**Stage 7: Destruction -** Consent data may be destroyed as per prevailing legislative needs.

*D. Blockchain for Consent Management*

Blockchain by its inherent design elements like decentralization, distributed peer-to-peer (P2P) network and implementation of an immutable ledger – enforces trust.

The following key characteristics of a Blockchain makes it suitable for Consent Management.

- **Distributed -** All transactions (monetary and non-monetary) that is included in a block is shared and updated across all nodes of the blockchain ledger network.

- **Secure -** Security is enforced through various cryptographic functions.

- **Transparent -** As all nodes and miners can access all the transactions on the chain, thereby enabling complete transparency on the blockchain.

- **Consensus Based -** All participants in the network must agree to validate a transaction using consensus protocols, thereby eliminating any monopoly. As more participants join a network the robustness continues to increase.

- **Flexible -** Event or condition-satisfiability based executions of custom codes (Smart contracts on Ethereum Virtual Machine (EVM) or Chaincode on Hyperledger Fabric) allows for flexibility of employing various logics, including Consent lifecycle management. Smart Contracts are self-verifying, self-enforcing and tamperproof.

*E. OConsent[xxvii] – Open Consent Protocol and Platform*

In this foregoing paper I intend to present a new protocol, and a platform architecture implementing the protocol - built on top of permissionless[xxviii] blockchain technology that can transparently address the Data Privacy and Consent Management concerns of digital businesses and legislators.

The platform intends to provide a transparent and non-repudiable protocol for full lifecycle management of consent for the end users as well as business and organization. The platform would also provide an audit track for the consent usage as per the agreed norms between end users and organization.

In short, the platform intends to empower the end users to make informed decisions and provide full control of their consent; and enable businesses to use such consent with confidence and in compliance of the prevailing legislations.

## F. Related Work

As our digital footprint has multiplied manifold over the past decades, and with organizations widely adopting the use of such personal data - there has been a growing acknowledgement that better data management practices must be devised, so that the control of one's own personal data remains with the data subject. Furthermore, with the wider adoption of Machine Learning and Artificial Intelligence among business there has been a surge in the demand for data collection for behavioral analytics. As discussed earlier, multiple legislations across the world are now trying to define standards around managing user's personal data and the necessary consent for its use, e.g., EU's GDPR. Consequently, there has been significant research and design of solutions that allow consent management recently.

One of the first uses of embedding attribution data onto blockchain was by the Blockstack domain name registration service. It used a Distributed Hash Table on a virtual crossover chain that separated the storage and blockchain operations. It stored the hashed key value pairs relating to the ownership and domain name details on the blockchain.

In 2018, Wang, Zhang and Zhang proposed[xxix] an access control mechanism with Ethereum, for managing entitlements of the files in the distributed Inter Planetary File System[xxx]. It employed a fine-grained customized attribute-based encryption. The keys for the attributes were generated and maintained by the data owner and disseminated to requesters.

The framework ADvoCATE[xxxi] proposed the use of Blockchain and smart contracts for managing consent and preferences for IoT devices. ADvoCATE extends the concept from the 2018 paper by the same authors[xxxii]. ADvoCATE uses Smart Contracts for directly embedding consents onto Ethereum public blockchain. Admittedly, this is not a cost-efficient solution as the price of Ether continues to rise. The paper uses XACML (eXtensible Access Control Markup Language)[xxxiii] based markup language as a standard policy language. XACML has had a mixed adoption in the industry[xxxiv]. There have been multiple improved[xxxv] markup languages to XACML, e.g., Policy Machine (PM)[xxxvi] based New Generation Access Control (NGAC). NGAC computes decision through a linear algorithm over non-conflicting policies, thereby making it operationally efficient over XACML that requires collecting attributes and running computations (matching conditions, rules and conflict resolutions) across a minimum of two different data stores - leading to extended complex computation steps. The proposed OConsent platform recognizes the clear advantages of NGAC over XACML and hence uses NGAC based markups to handle incoming consent and data access requests from Data Controllers. One of the key components of ADvoCATE is the Intelligence Component, that uses Fuzzy Cognitive Maps[xxxvii] (FCMs) to resolve conflicting policies for access requests. Fuzzy Cognitive Maps are popular for modeling complex systems but are known to be plagued by time lags between causes and observed effects. Consequently, Generalized Fuzzy Cognitive Maps[xxxviii] (GFCM) and Generalized Rules Fuzzy Cognitive Maps[xxxix] (GRFCM) have been recently proposed to overcome such challenges. ADvoCATE also proposes a recommendation module, based on Cognitive Filtering for recommending personalized rules.

Consentio[xl] is another platform that looks to address the management of consent with blockchain. Consentio uses Hyperledger, which is a permissioned blockchain. Hyperledger Fabric[xli] is known to be faster while processing transactions when compared to the Permissionless blockchains like Bitcoin and Ethereum. However, having a permissioned blockchain inhibits the wider adoption of the platform, and arguably is against the inherent idea of a decentralized blockchain – where the admission on the platform is tightly governed. The platform also maintains a World State Store – which is a key value store maintained by Hyperledger Fabric, with simplistic GET and PUT requests. This provides a high throughput over and above the conventional Hyperledger Fabric's gains. OConsent uses an Open Source Distributed In-Memory Key Value Store - Apache Ignite[xlii], that provides extremely fast Global State Store for the platform with simple PUT/GET requests as well as fully compliant ANSI SQL interface with strict transactions and complex analytical querying needs. Consentio does not propose any standardize markup languages for access control policies.

Truong, Sun, Lee and Guo proposed[xliii] to use a permissioned blockchain based on Hyperledger Fabric for consent management and provenance, similar to Consentio. Consequently, although the platform produces a higher throughput as exhibited by the benchmarks in the paper – it may not be widely adopted unlike Bitcoin and Ethereum. It must be noted that, similar throughputs[xliv] are possible using Sidechains, State-channels and Plasma – that OConsent uses. The platform uses access tokens and log ledgers for controlling access and tracking usage. It uses MongoDB as a backend for its profile management webservice. The platform uses the built-in Hyperledger Fabric ordering service with Apache Kafka. The platform does not account for any anonymity or pseudo anonymity concerns.

In the studies encountered, many designs included either using direct public blockchains for embedding consent hashes or using a permissioned Hyperledger Fabric based blockchain. Both of these approaches have limitations, in throughput and adoption, respectively. OConsent proposes a mixed approach, using a Local Ethereum based sidechain for granular embedding of consent hashes and versioning, while using a combination of Ethereum Main net and Bitcoin Main net for capturing the state of the local platform. This approach enables a high likelihood of adoption as the local chain is Permissionless and guaranteed high throughput as it uses a Sidechain. OConsent also explores the use of State Channels and Plasma based 2nd Layer Scaling. An In-memory Distributed Global State Store also forms a key component of the platform enabling high throughput and low latency.

None of the explored platform offers any anonymity or pseudo anonymity options. OConsent provide Surrogate[xlv]

ID and Zk-SNARKs[xlvi] based zero-knowledge proof for anonymity needs. Another key feature that is only attributed to OConsent platform is the embedding of the Trusted Timestamps Proofs. Provable and trusted timestamping is important as it enables non-repudiable assertions that a consent was generated at a particular point-in-time. This is vital as we move towards increasingly real time interactions and consequently, must cater time-exactness of a consent availability or revocation. OConsent also includes Time Leasing of Consent – which is a powerful option to make sure consents are not awarded perpetually and that expirations can be enforced.

## II. Platform Design and Architecture

### A. *Requirements and Design Considerations*

The following section discusses the various functional and non-functional requirements as well as design considerations.

*1) Key Functional Requirements for Consent Management Platform*

- **Freely given:** Consent must be provided by the Data Subject (DS) freely and completely optionally without any coercion.

- **Informed, Granular and separate:** Purpose for which a consent is sought must be clear, atomic, and definitive. Separate consent must be sought for separate scope. A consent requirement and context must be concise and specific. E.g., a consent pursued for marketing must not automatically be reused for analytics.

- **Unambiguous grant:** It must be clearly demonstrated that an individual (Data Subject) has granted consent. There must not be any ambiguity on the affirmative action.

- **Named:** Consent agreement must clearly define the Data Controller and Data Processing organization and any third parties involved. The platform must establish and manage verifiable identities.

- **Avoid default opt-in:** The Data Controller or consent seeker must avoid using prefilled checkboxes or forms for seeking consent. The Data Subject must explicitly demonstrate affirmative opt-in actions.

- **Right to withdraw consent:** End Users (Data Subjects) must be clearly notified at the time of obtaining consent that they may revoke consent any time, and that there will not be any residual consent-based actions after the withdrawal.

- **Regular Review:** Consent validity and usage must be continually reviewed. A consent management platform must thereby account for scheduled checks. The platform must also allow 3rd party auditors and reviewers to validate such consent usage claims.

- **Time based lease[3]:** Consent granted must not be indefinite, and should include some time bound default expiry, if not explicitly overridden. This also requires that a platform must ensure a trusted timestamp, so that time-based validity may be enforced.

- **Right to Forget:** The End User may choose to exercise his/her Right to Forget, which would entail a complete destruction of stored personal data from the platform and/or the Data Controller and Data Processor.

*2) Key Non-Functional Requirements for the Consent Management Platform*

- **Security**
    - **Confidentiality and Privacy:** The platform must ensure that necessary controls are included so that confidentiality and privacy is maintained for all stakeholders. These may include segregating roles and actions within the platform as well as segregation of duties among the platform and blockchain node operators. The platform must operate with the notion of least-privileges.

    - **Anonymity:** The platform should provide options to Data Subjects (end-users) to operate with necessary anonymity when desired.

    - **Non-Repudiation:** Trust in the platform can be established only when it operates transparently, and all actions are supported by verifiable proofs. These proofs include verification of digital signatures, timestamping, fingerprinting, etc.

- **Performance:**
    - **Latency:** The platform must operate with low latency for most of the processes and associated actions. As real time processing needs become center stage – its paramount that the platform should be able to support actions like consent querying and consent revocations actions withing a few seconds. This may entail employing distributed in-memory cache for Consent Queries responses and Circuit Breakers for immediate consent revocation.

    - **Throughput:** The platform must be able to handle high throughputs, order of at least 500 tps[xlvii] (transactions per second)

[3] OConsent includes a Time Lease based smart contract implementation that expires the Consent Lease after an agreed time.

as is demonstrated by contemporary implementation.

  - **Scalability:** The platform should be ideally horizontally and linearly scalable, i.e., it should be able to support higher workloads with constant performance with the addition of new hardware. A microservices based architecture must be embraced for granular scalability.

- **Reliability:** The platform should be able to operate reliably with a reasonably expected performance.

- **Availability:** The platform should be fault tolerant and must continue to operate even if there are node failures and network partitioning.

- **Modifiability:** As the platform will continue to evolve, it must support extensibility and modifiability. These would require that Smart Contracts must be properly versioned and designed so that newer and latest versioned Smart Contracts can be deployed without breaking changes. All interfaces and APIs must support extensibility for integrating with 3rd Party Service Providers.

- **Maintainability:** The platform should be easy to maintain, i.e., installing upgrades and patches, without extensive downtimes.

- **Usability:** Providing a simple, consistent, and engaging UI/UX is key to attracting and retaining Users.

- **Cost:** The design should cater for reducing operational cost. Infrastructure should be based on commodity non-specialized hardware. Where applicable, Open-Source tools and frameworks should be adopted. Special attention must be given to reduce the transaction cost on the Global Public Blockchain, e.g., Ethereum and Bitcoin. This may entail deciding on the right batch size to include for fingerprinting on Bitcoin/Ethereum.

### B. *Logical Architecture*

Refer Figure 3.

#### 1) *Key Terms*

Following are the key terms that is used throughout this treatise.

- **Consent Agreement**
  Contract that lists all the details of a consent, e.g., parties involved – data subjects, data controllers, time validity of the consent, context/purpose of the consent.

- **Consent Proof**
  Consent Proof is a collection of cryptographic proofs that guarantees the non-repudiation of the Consent Agreement. Proofs include provable timestamp, snapshot fingerprinting and/or full Consent Agreement's hash sum fingerprinting, consent versions lineage, etc. This is a JSON-LD[xlviii] document.

- **File Hash**
  A fixed length string that is the output of passing a file's content through a hashing function, e.g., SHA 256. Every file with a different content produces a different hash value, whereas a file with same content will produce the same hash value. The Hash value generated is thus essentially the fingerprint or identity of the file and its contents.

- **Signature**
  A file may be signed with a Private Key, to establish the ownership of the key and to prove that a file has not been modified. A signature is usually a fixed length string of characters. A user's Private Key is used to sign a file, whereas its Public Key is used to verify the ownership of the file.

- **Data Access Key (DAK)**
  Data Access Key is used to access the Data Subject's (End User's) Data stored external to the OConsent Platform, after the Data Controller (or data requester) has proven that he/she has the necessary consent permissions to access such data.

#### 2) *Key Actors*

- **Data Subject (DS)[xlix]**
  Data Subjects (or DS) are End Users who provide consent. DS are the primary actors on the platform and have full control on the consent lifecycle from creation to usage to deletion.
  DS interact with the platform through Mobile Apps available on iOS and Android Devices. The Apps serve as one-stop source for information on all active consents as well as well as their management.

- **Data Controller (DC)[l]**
  Data Controllers (or DC) are the Business or Organizations that seek consent from the end users. DCs interact with the platform through a web portal. Every DC has tiered accounts – starting with the primary admin account, followed by other secondary accounts with various permissions. These secondary accounts may have varying access rights based on the different business units they belong too. An example of the tiered account would be a Bank – that has one primary account with super privileges, while multiple secondary accounts for Consumer Banking, Institutional Banking, and Digital Banking.

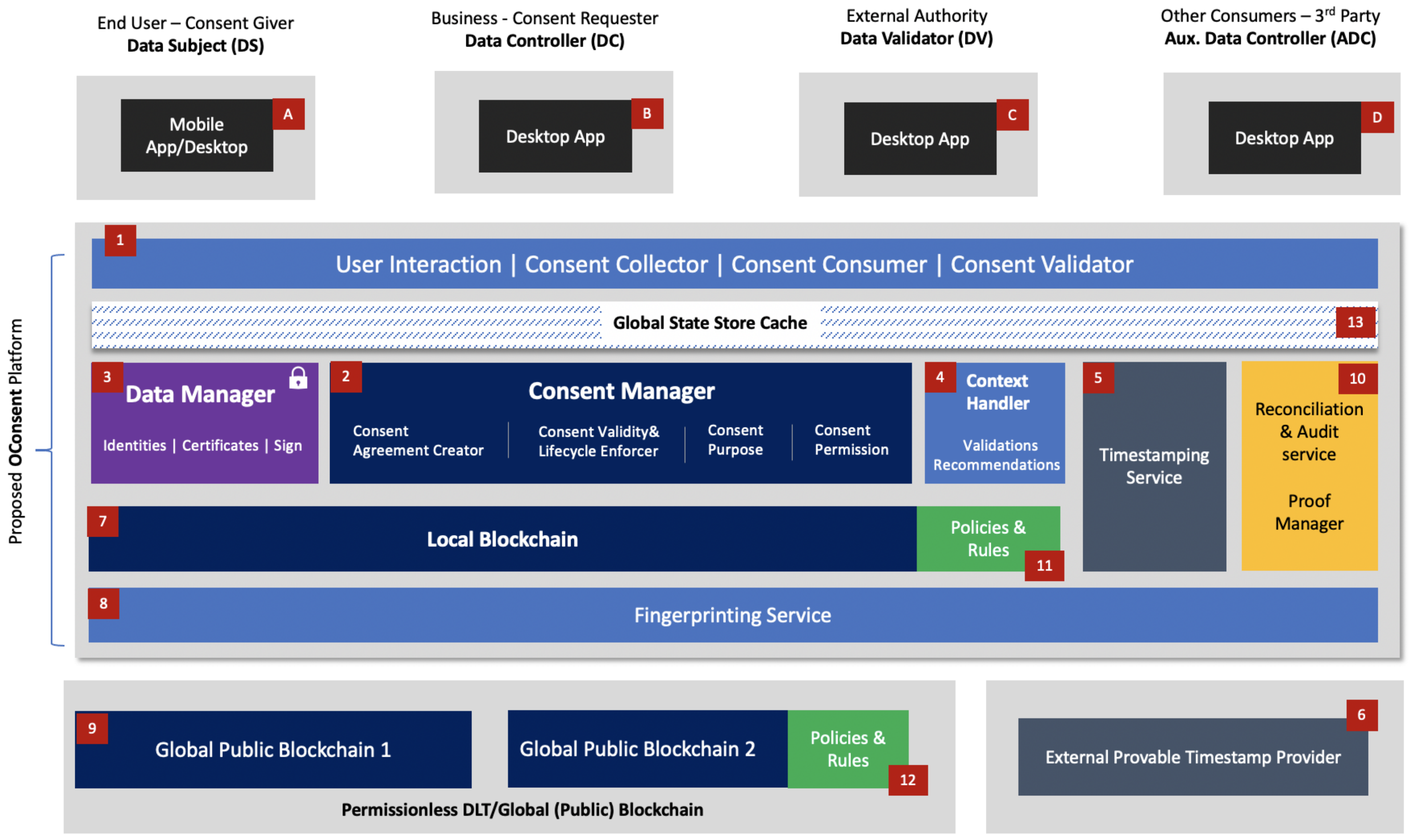

*Figure 3. Logical Architecture of OConsent Platform. The above figure represents the various logical components that make up the OConsent Platform. The subsequent sections discuss the platform in detail*

- **Data Validator (DV)**
  Data Validators (or DV) are independent actors who may validate if an organization or business is using a DS's consent in accordance with the DS's permission. Typically, DVs can be external auditors (both governmental as well as non-governmental). DVs requests for consent validations and proofs are served through the immutable fingerprints on the public blockchains.

- **Auxiliary Data Controller (ADC)**
  Auxiliary Data Controllers (or ADC) are third party entities that may inherit consents from Data Controllers (DCs). Propagation of consent via DCs must be in-accordance with the Data Subjects (DS) explicit permission and must not be assumed. ADCs are typically DC business partners. Before a consent is federated or propagated to ADCs it undergoes validations for rules conflicts.

- **Other Actors (OA)**
  i. Platform Operators (PO)
  ii. Local Blockchain Miners/Participants (LB):
  These are users who operate an instance of the Local OConsent Blockchain Node. These may other DS, DC, DV or ADCs.
  iii. Global (Public) Blockchain Miners (GB):
  These are miners from the public who may or may not be participating in the OConsent Platform.

*Other Actors (OA) do not have direct access to Personally Identifiable Information (PII) and the platform operates strictly on the principal of least privilege. Do note that, PII stored data and its Hash are decoupled, and that only the hashed identities of the datasets are fingerprinted.*

*3) Key Components*

The following section describes the key components of the OConsent Platform. Please refer the labels against the various components in Figure 3. Logical Architecture of OConsent Platform above.

- **Interactions Layer**
  This is the interface layer with which the various actors interact with the platform. This is also the interface that users use to capture and manage their consent and data. This layer provides the full suite of actions governing the consent lifecycle from definition and enforcement of data rights, data erasure and data/consent modification.

- **Consent Manager**
  Consent Manager is the heart of the platform and undertakes multiple functions, Consent Agreement Creator, Consent Validity and Lifecycle Enforcer, Consent Purpose and Consent Permissions. The Consent Manager takes in the "Consent Request" from the Data Controller and the "Expression of Consent" from the Data Subject and enforces that the data is handled according to the consent terms and privacy statutes. The Consent Manager also maintains multiple versions of the consent for audit and tracking purposes. Only the current (latest) version of the consent is enforced. It also coordinates with other modules to trigger the metadata captures associated with the consent lifecycle, e.g., who created the consent, for whom was the consent created, the unique hash associated with the consent, timestamping requirements, data vaulting, etc. The consent manager is responsible for maintaining the "Consent Proof".

- **Data Manager**
  Data Manager is responsible for securely storing various data and only allow authorized access. The data types include, User's PII and non-PII Attribute Lists, Consent Metadata, Surrogate IDs, signature keys.

  *Note that the platform does not physically store the Data Subject's data. Only the column metadata is retained. The Data Subject (DS) is responsible for storing and maintaining his/her data off-OConsent Platform either on AWS S3, GCP GCS, Azure Blob Store, Storj or some other decentralized store. DS or the platform that hosts DS's data must release the data only after Data Controller's demonstrated proof, e.g., the Data Access Key (DAK)*

- **Context Handler**
  Every action performed by the actors on the platform have an associated context. The Context Handler is a reactive service responsible for interpreting the context and triggering a relevant action. For example: Data Subject (DS) may respond to a consent request from a Data Controller (DC). A context handler provides the following key functionalities:
  i. Logically validate the context and associated rules for correctness or conflict.
  ii. Trigger Policies and Rules.
  iii. Recommend rules associated with the contexts (e.g., recommend rules of Consent Agreement based on the Data Controller's domain – ecommerce site)

- **Timestamping Service**
  This service invokes the External Timestamp Providers and embeds the timestamps into the generated Consent Proof.

- **External Provable Timestamp Provider**
  Multiple external Timestamp providers may be used to prove that an action ("Consent

Agreement") happened after a certain point in time. This ensures the non-repudiation of the Consent Agreement.

- **Local Blockchain**
  A local blockchain is maintained to capture the Consent Agreement and Consent Proof details. It also embeds Smart Contracts/Chaincode that are executed in response to various events incoming from the Context Handler. The blockchain is generally a compatible local implementation of a public global blockchain, like Ethereum. Miners of the local blockchain include multiple DCs, DVs and ADCs. A DS may also choose to participate in the local blockchain by running a node.

- **Fingerprinting Service**
  This service takes the snapshot of the local blockchain and periodically publishes it to the global public blockchains like Ethereum and Bitcoin. The fingerprinting service may be scheduled by time or by volume of processed Consent Agreements.

- **Global Public Blockchain**
  These are public blockchains e.g., Bitcoin and Ethereum.

- **Reconciliation and Proof Manager**
  This service provides the necessary cryptographic and finite proofs for Data Validators. Proofs contain the validity of a Consent Agreement and its current usage as per the stipulations in the agreement.

- **Policies and Rules (Local Blockchain)**
  These are policies that are executed automatically based on the incoming legally relevant events and actions according to the terms of the Consent Agreement. These rules modify the state of Consent Proofs on the Local Blockchain of the platform.

- **Policies and Rules (Global Blockchain)**
  Similar to policies and rules for Local Blockchain.

- **Global State Store Cache**
  This is used to increase throughput and reduce the latency of the platform. It maintains a key-value store of Data Subject and Data Controller's agreement state in memory for fast retrievals. All front facing API requests are also served through the cache where applicable.

*C. Technical Architecture*

Refer Figure 4

The OConsent Platform is composed of multiple containerized microservices orchestrated by Kubernetes or OpenShift. These services are fault-tolerant and independently scalable running on containerized infrastructure. The Local Blockchain although shown as single instance as part of the architecture is operated across multiple distributed nodes, that are owned by independent parties.

Following are the components of the platform.

- **API Gateway**
  API Gateway serves a collection of APIs to the frontend mobile and web applications. API categories include for platform authentication, consent management, data authorizations, data uploads, data management, reporting, etc. The gateway also has a HTTP layer 7 load balancer that routes the frontend traffic to multiple instances of the backend microservices.

  APIs follow the standard OAUTH 2.0 based authentication for HTTP REST based service transactions.

- **Policy Markup Parser**
  The API payloads for Consent request and data access control follow the standard Next Generation Access Control (NGAC)[li] / Policy Machine[lii] (PM) formats. All these formats define a declarative fine-grained attribute-based access control; however, PM/NGAC is the newest and most versatile. NGAC is discussed in the subsequent chapters.

- **Consent Manager**
  The consent manager is a logical composition of multiple services. These services are:

  i. **Consent Agreement Creator (CAC)**
     This service is responsible for coordinating the agreement creation process and handles tasks like generating the Seed Agreement ID, the JSON Agreement Template and the Signed Agreement Hash.

     The service fetches pre-registered identities (public and private keys) of the Consent Subject and Consent Controller from the Identity Store. With the fetch keys it double signs the Consent Agreement to generate the final signed agreement hash. It then emits, the event for "timestamping" the agreement from the Trusted Timestamp Provider.

     Once the Signed Consent Agreement has been timestamped, this service triggers the Local Blockchain Smart Contract for embedding the agreement hash onto the chain. It also periodically takes Hash of the Hashes on the Local Blockchain and fingerprints onto the Global Public Blockchains, Bitcoin and Ethereum.

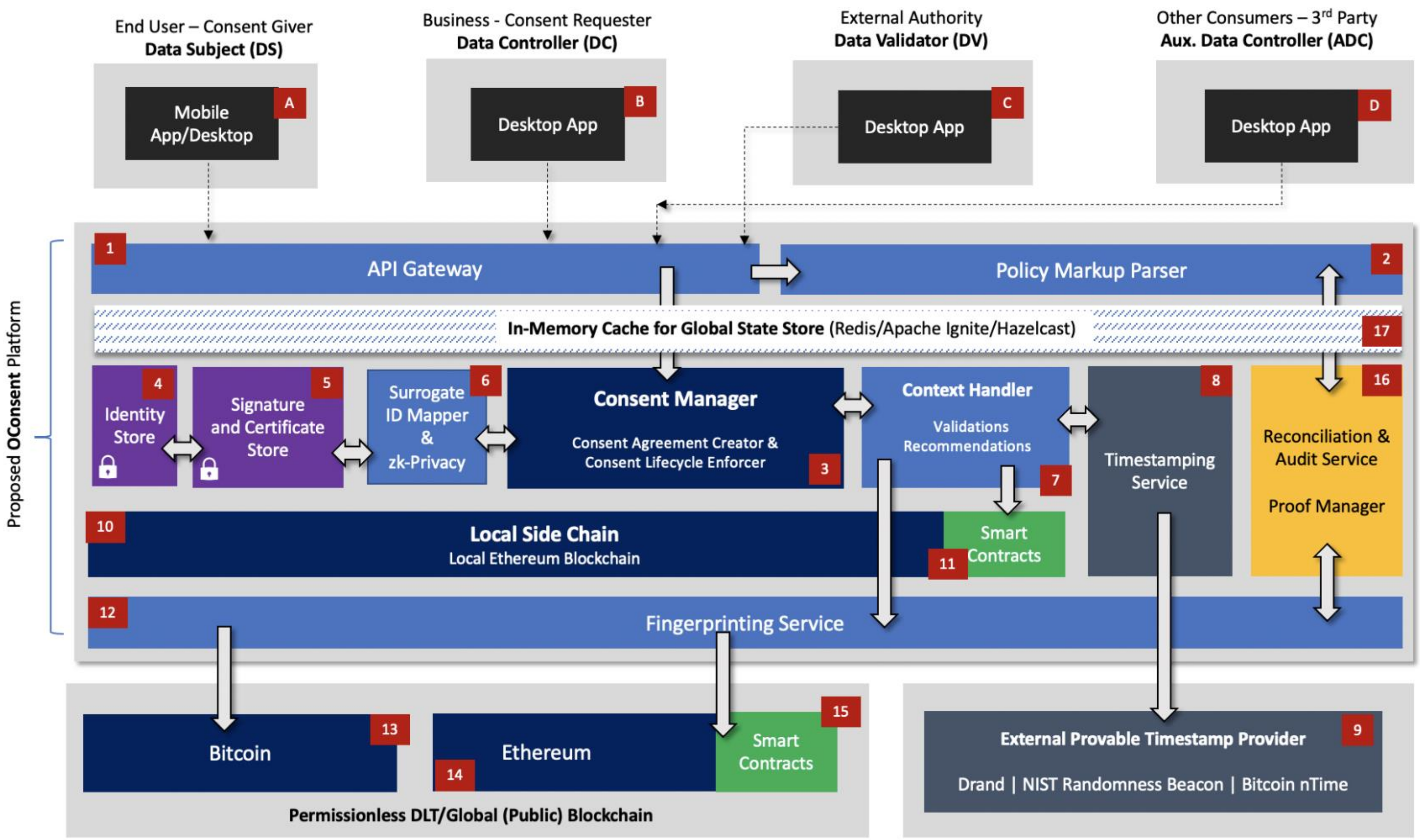

*Figure 4. Technical architecture of the OConsent Platform*

ii. **Consent Validity and Lifecycle Enforcer**
This service is responsible for end-to-end lifecycle management and enforces the validity of the consent agreement on the local blockchain through the smart contracts. If an agreement has been updated, it triggers a smart contract for setting up the new version of the Smart Contract.

- **Identity Store**
  Stores the various mapped Primary Identities and Surrogate IDs per Data Subject (DS) and Primary identities for all other actors. This store also saves Government issues Official Digital identities as well as Offline non-digital Identities.

- **Signature and Certificate Store**
  Store for Public Keys (and Private Keys, if uploaded by DSs, DCs, ADCs). Multiple keys are stored per identities and rotated for greater security.

- **Surrogate ID Mapper and Zero-Knowledge (Zk) Privacy**
  Module that generates Surrogate IDs and generates ZK-SNARKs based privacy proofs. Please read Section Error! Reference source not found. for more.

- **Context Handler**
  Service that analyses the context for which a Consent is sought and checks if it's a valid context – if yes, it proceeds to generate a pre-defined template for consent agreement. If, however, a Consent Agreement exists between the DS and DC, and a new context is encountered – it either fails or prompts parties for an addendum contract.

- **Timestamping Service**
  This service is responsible for contacting the trusted Timestamping Authority (TSA) for generating a timestamp proof for embedding into Contract Agreement. Please read Section Error! Reference source not found. for more.

- **External Provable Timestamp Provider**
  An external authoritative service that generates timestamp or beacons periodically, and which can be leveraged to prove a document existed after a certain point in time. Please read Section Error! Reference source not found. for more.

- **Local Blockchain**
  The OConsent platform employs a Sidechain as its Local Blockchain, for storing consent agreement hashes, proofs and running smart contracts. The sidechain in case of OConsent is an locally running Ethereum Blockchain. This Ethereum blockchain is completely independent of the Global Main Ethereum Network. Please refer to Section Error! Reference source not found. for more.

- **Smart Contract on the Local Blockchain**
  Smart Contracts that carry out various workflows and key functionalities of the Platform. Smart Contracts have key features like self-verifying, self-enforcing and tamper-proof that make them extremely desirable components of the platform. Please refer Section Error! Reference source not found. for more.

- **Fingerprinting Service**
  From time to time, the snapshot hashes of the Local Blockchain are inserted into the Global Blockchains (Ethereum and Bitcoin). This is done so that a publicly verifiable proof exists, and in the event bad actors of the Local Sidechain corrupt the same should fail and fallback to the snapshot on the Main Global Blockchain.

- **Bitcoin Network**
  The official public Bitcoin Network

- **Ethereum Main Network**
  The official public Ethereum Main Net

- **Smart Contract on Ethereum Main Network**
  Smart Contract for downloading "Consent Proofs" from the Main Network.

- **Reconciliation and Proof Manager**
  Service that runs reconciliation reports and handles on-demand proof requests by Data Controllers or Data Validators (external auditors).

- **In-Memory Cache for Global State Store**
  An Apache Ignite based distributed key value store that provides a layer of acceleration for the OConsent Platform. It stores, key-value pairs of Data Subject and Data Controllers, as well as cached API results. Apache Ignite is an open-source product and provides reliable, ACID transactions, ANSI SQL compliant – in memory data grid

### *D. Consent Management Process*

The following sections describe the essential workflows of the platform

#### *1) Consent Agreement Creation Activity Diagram*
Refer Figure 5.

#### *2) Data Access Activity Diagram*
Refer Figure 6.

#### *3) Consent Agreement Proof Generation Process*
Refer Figure 7.

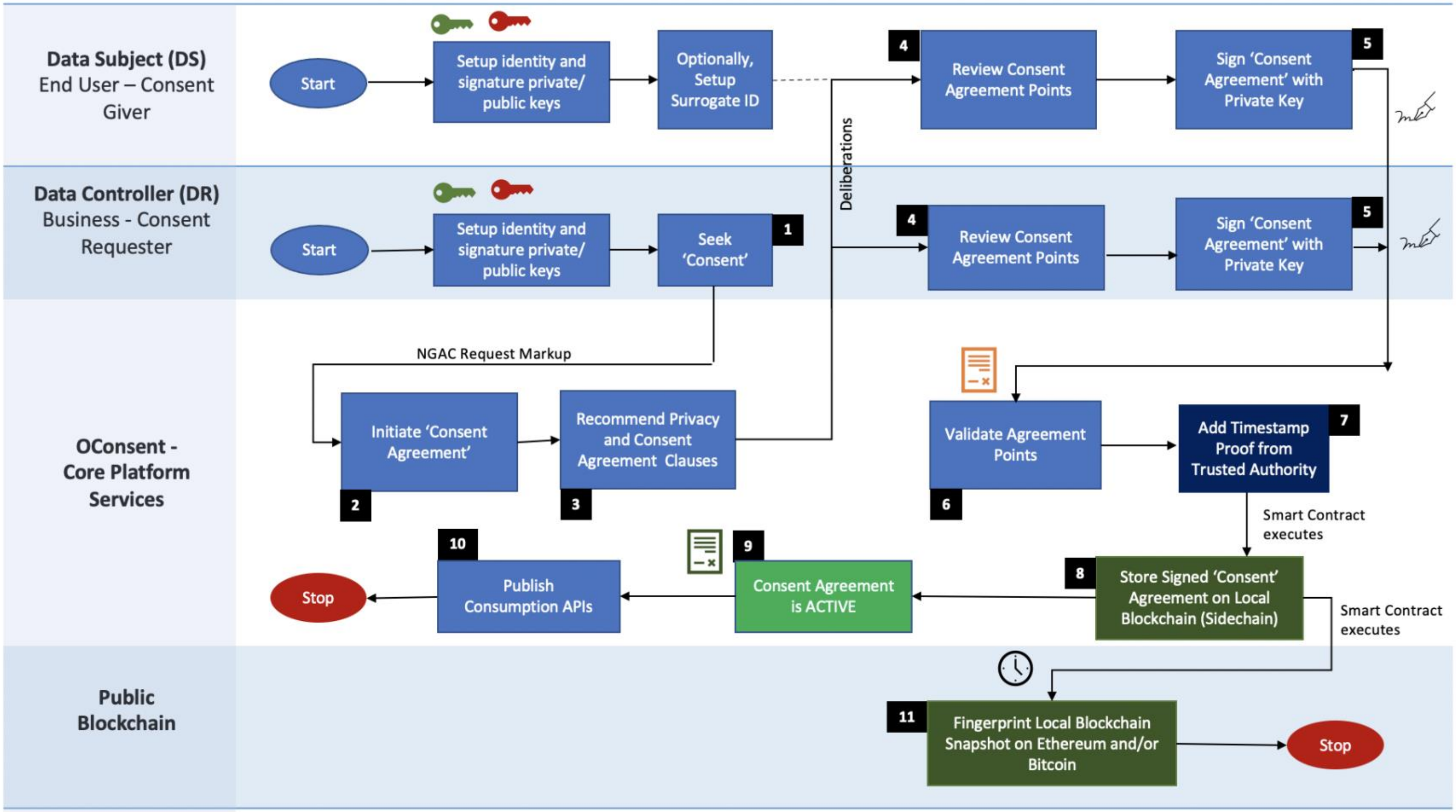

*Figure 5. Process flow and interactions during the Consent Agreement creation phase. Please refer to Data Access Activity Diagram for the logical next steps on how Data Controller accesses Data Subject's data.*

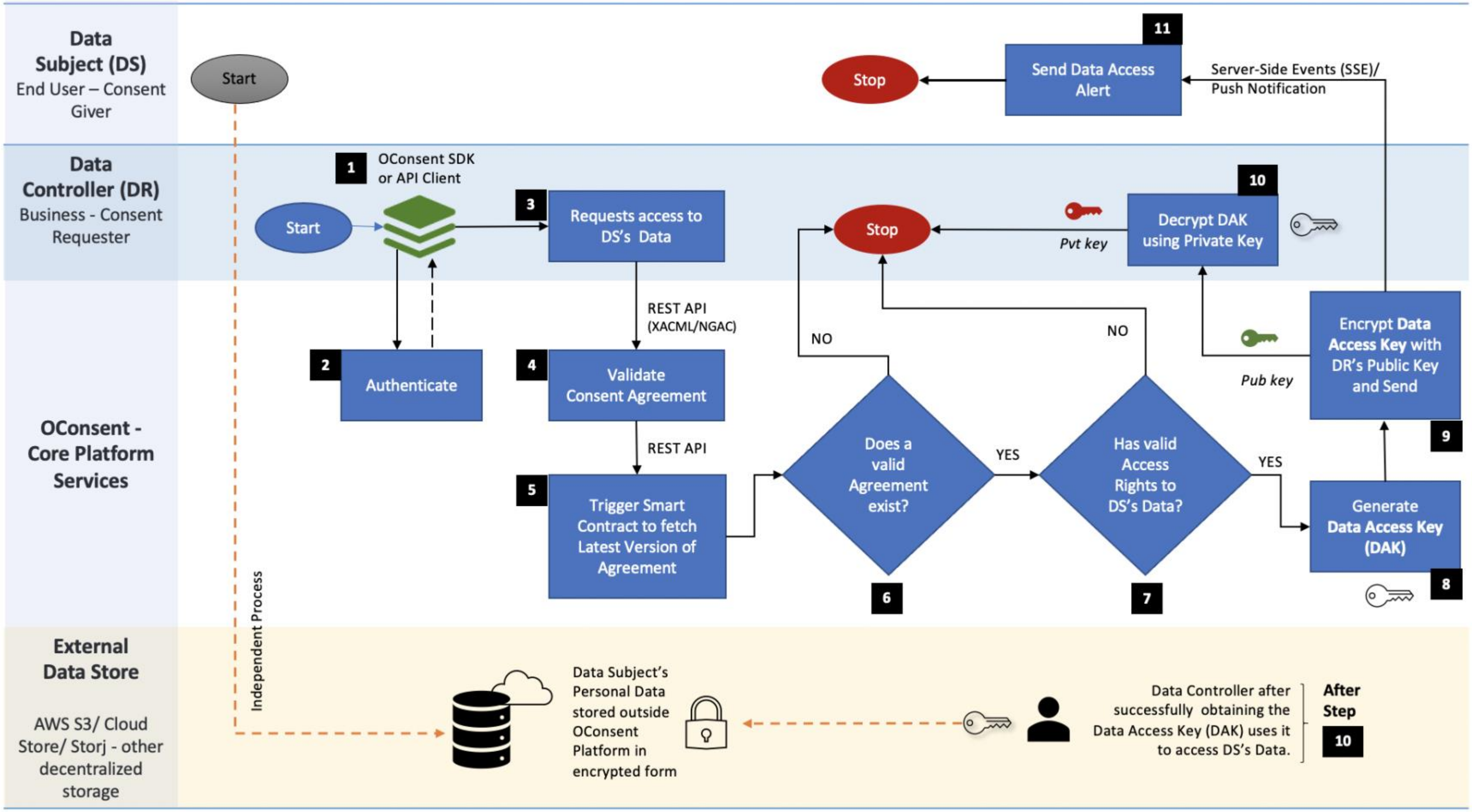

*Figure 6. How Data is accessed based on the validity of the Consent Agreement.*

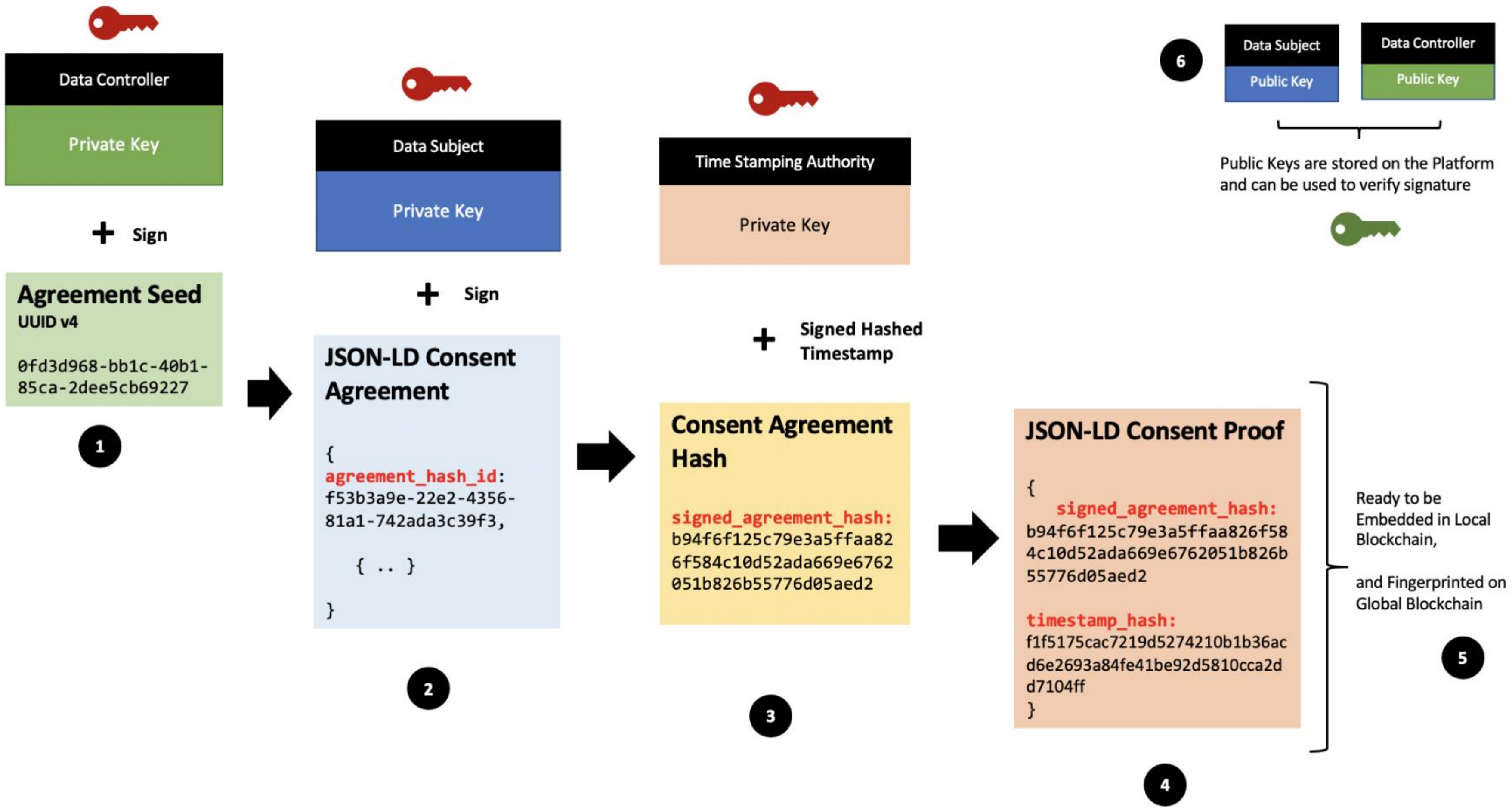

*Figure 7. The stepwise process of Consent Agreement Proof generation*

*a) The process to generate a keypair for signing*

```
openssl req -nodes -x509 -sha256 -
newkey rsa:4096 -keyout "$(whoami)s
Sign Key.key" -out "$(whoami)s Sign
Key.crt" -days 365 -subj
"/C=IN/ST=Noida/L=NCR/O=OConsent
Network/OU=IT Dept/CJ=$(whoami)s Sign
Key"
```

*b) The process to sign the document with Private Key and OpenSSL*

```
openssl dgst -sha256 -sign "$(whoami)s
Sign Key.key" -out consent.json.sha256
consent.json
```

The output of the above will result in the signed file consent.json.sha256 with hashed content of the file consent.json. Where consent.json is the JSON-LD Consent Agreement as referred in Section **JSON-LD Consent Agreement Structure** . The platform would generate more relevant names for the agreement and may not be as simplistic.

*c) The signed file can be validated as below*

```
openssl dgst -sha256 -verify <(openssl
x509 -in "$(whoami)s Sign Key.crt"  -
pubkey -noout) -signature
consent.json.sha256 consent.json
```

*d) Process Flow*

- **Consent Agreement Seed Generation**
  A UUID v4 based seed is generated and Signed with the Data Controller's private key. This signifies the initiation of Consent Request and establishes that the request for consent indeed came for the specific DC.

- **Consent Agreement Hash ID Generation**
  The agreement hash ID is generated as an output of Step 1 above, and is included as part of the Consent Agreement JSON-LD.

- **Signed Agreement Hash Generation**
  Once the Consent context and scope is agreed, the agreement hash ID is signed by Data Subject using his/her own private key to generate the Signed Agreement Hash.

- **Timestamp Hash Generation**
  The generated timestamp hash from the Timestamping Authority is included in the Consent Proof JSON-LD.

- **Consent Proof Generation and Fingerprinting**
  Once the Consent Proof is ready, it may be further signed with the OConsent Platform's Private Key and embedded onto the Local Blockchain first, followed by the Global Blockchain - Ethereum and Bitcoin.

- **Provability**
  The hashes and signatures generated through each step may be proven by applying the Public Key of the actors sequentially.

*e) JSON-LD Consent Agreement Structure*

Refer Figure 8

*f) JSON-LD Consent Proofs*

Refer Figure 9

### *E. Trusted and Provable Timestamping*

Trusted Timestamping helps to track when a Consent Agreement was created, modified, or cancelled. Trusted Timestamping authorities provide the necessary cryptographic proof that makes repudiation of a consent event on OConsent Platform highly unlikely. None of the related work in blockchain based consent management employ a Trusted Timestamp Anchor. OConsent is the first Platform/Protocol that leverages provable timestamping for point-in-time validations. It is extremely useful for purposes of administration and audit. As the timestamp proofs can be publicly validated, the stampers integrity is unrepudiated.

*1) Timestamp Generation*

Refer Figure 10

*2) Checking the Timestamp*

Refer Figure 11

```
{
      @context: https://w3id.org/oconsent/v1
      type: “OConsent – Open Consent Agreement”,
      agreement_hash_id: f53b3a9e-22e2-4356-81a1-742ada3c39f3,
      agreement_version: 1.01,
      linked_agreement_hash_id: 365b5f44-cac8-4e78-8bfa-8d07899c6385

      metadata: {
        data_subject: {
            name: “Mr. XYZ”,
            id: 7a2a83b1694940f38d6a2a8f50e4d979
                },

        data_controller: {
            name: “ABC LLC.”,
            id: 478ecb5f2b674ad18976007d64c069de
                },

       data_controller_aux:{..}

        agreement_date: 02/09/2020
        is_transferrable: FALSE

      },
      consent_scope: [
        “marketing”: {
              data_attributes: [datasetA:attr1, datasetB:attr2],
              expiry: 01/12/2020
             },

        “analytics”: {
              data_attributes: [datasetB:attr2, datasetZ:attr4],
              expiry: 01/11/2020
             }
      ],

      monetization_enabled: FALSE
      monetization_scope: { .. }
}
```

*Figure 8. Various attributes of a sample JSON-LD OConsent Agreement*

```
{
      @context: https://w3id.org/oconsent/v1
      type: "OConsent – Open Consent Proof",

      agreement_hash_id: f53b3a9e-22e2-4356-81a1-742ada3c39f3,
      linked_agreement_hash_id: 365b5f44-cac8-4e78-8bfa-8d07899c6385

      signed_agreement_hash_id: b94f6f125c79e3a...d52a51b826b55776d05aed2

      timestamp_proofs: {
            nist_randomness_beacon:
                        9e82461318bd5f55282dbb05a1b29b15b381cf7d46a1d94addfc8ec4
                        17772b36751b57b653f7832205b3631fa427e61dd096fd24ae14f205
                        d5a47522785ce6a9,

            btc_ntime_hash:
                        0000000000000000000f716d50cc2f0434a620d82c7e2157e12f646
                        3a5bd487,

            drand_hash: {...}

      },

      URIs:[
            https://beacon.nist.gov/beacon/2.0/chain/1/pulse/1151283,
            https://www.blockchain.com/btc/block/0000000000000000000f716d50cc2f
            0434a620d82c7e2157e12f6463a5bd487,

            ...

      ]

}
```

*Figure 9. Sample JSON-LD OConsent Proof*

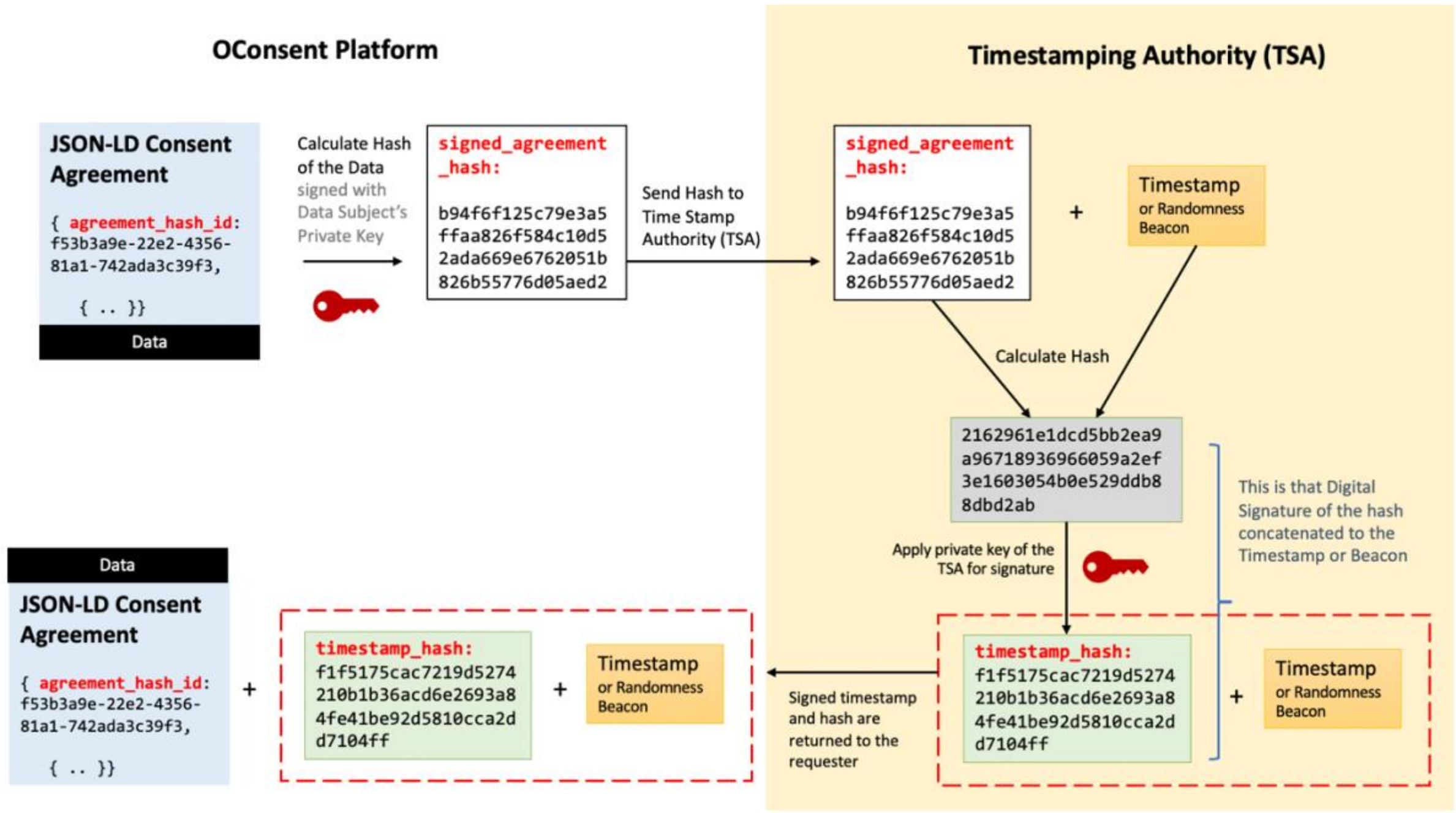

***Figure 10.*** *The process flow describing how the trusted timestamp hash is generated for any given Consent Agreement*

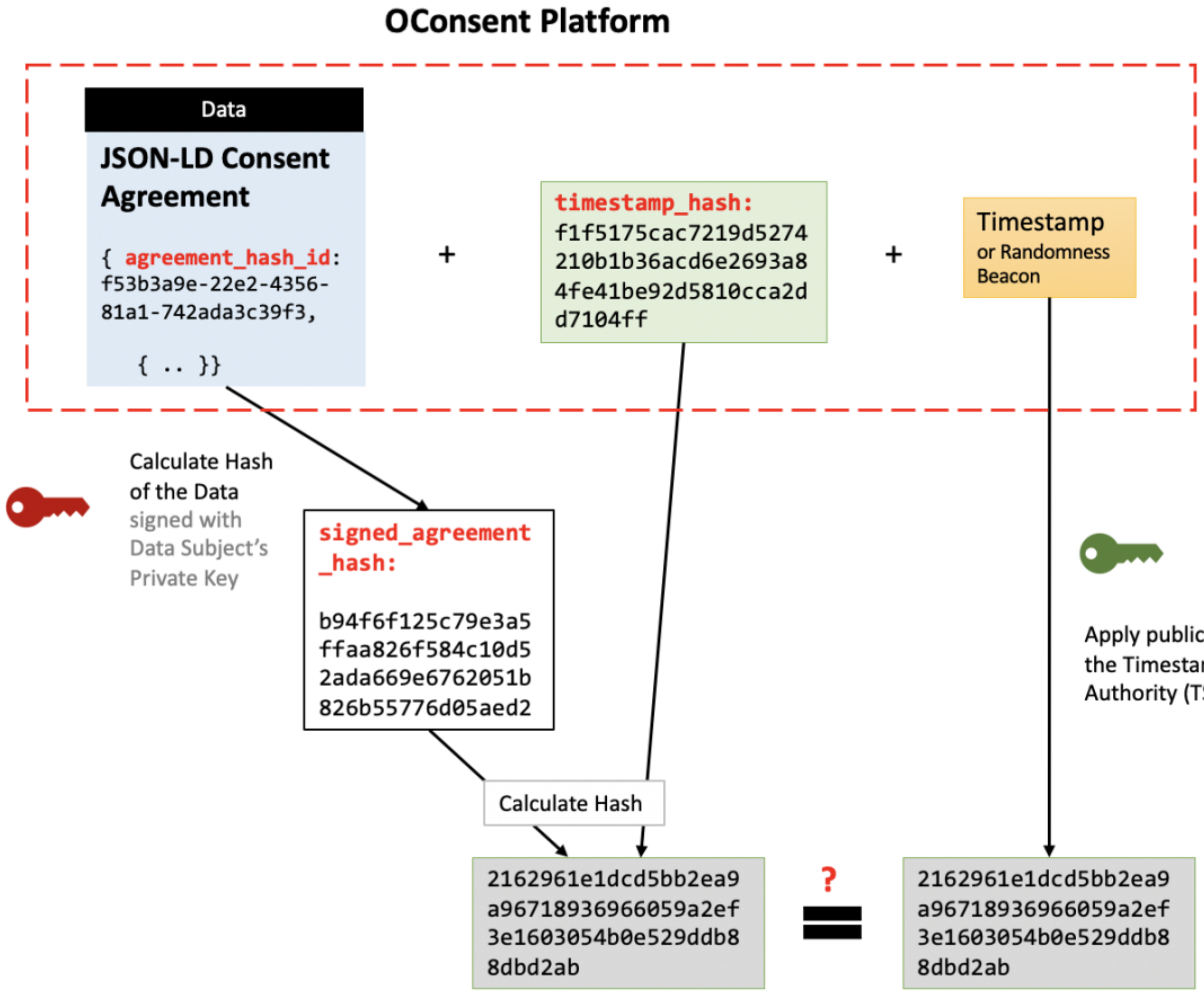

***Figure 11.*** *How the Trusted Timestamp may be validated.*

If the calculated hash code equals the result of the decrypted signature, neither the document or the timestamp was changed and the timestamp was issued by the Trusted Timestamp Provider (TTP). If the calculated hash does not match, either of the previous statements is not true.

1) *Trusted Timestamping Authority (TSA) Options*

*a) Using Bitcoin Header nTime*

The public Bitcoin blockchain's every block header has a field called 'nTime'[liii]. As part of the protocol, when a block is accepted by the network the header field 'nTime' must be updated with the approximate block creation time. This time has a variance of 2 hours (<1 day). This is not an accurate time reference, however if we are looking for an attestation that a given data/record existed at some point time after this 'nTime' – then this approach would suffice.

> Block 659792[liv] with an nTime of 2020-12-04 00:57 has a hash value of **0000000000000000000aa23344fcefaafa535d40f3f6185aa71c05f361a50067[4]**

*b) Using NIST Randomness Beacon*

The event of consent capture must timestamp - provably through an absolute (close to an absolute) clock. OConsent provides the option of using the National Institute of Standards and Technology (NIST) Beacon to prove the exact time when consent was capture/modified or revoked.

Every minute, the Randomness Beacon publishes a value created by a network of random number generators. Beacon values are generated by specialized hardware that uses quantum effects to generate a sequence of truly random values, guaranteed to be unpredictable, even if an attacker has access to the random source. These Beacon values can be use as Timestamp Identifiers and demonstrate that a certain event happened after (if not exactly) a certain clock "minute".

*c) Using Drand Randomness Beacon*

Drand [lv] is a distributed randomness beacon daemon. Multiple Drand running servers produce collective, publicly verifiable, unbiased, unpredictable random values at fixed intervals using bilinear pairings and threshold cryptography. Drand development is supported by Cloudflare[lvi].

Drand is arguably much better than NIST Randomness Beacon, as the latter hasn't had a steady development and release; and has been marred with reliability issue.

Drand also scores in terms of the redundancy, as it is a distributed service whereas NIST Random Beacon is run by a single operator.

Furthermore, Drand has a better coverage as it publishes beacons every 30 seconds, whereas NIST Random Beacon is only emitted every 1 min.

## F. OConsent Local (Sidechain) Blockchain

The local blockchain forms a key component of the OConsent Platform. It's on the local blockchain that the Consent Agreements are embedded, including the Timestamp proofs. The local blockchain also maintains all versions of a Consent Smart Contract. Refer to the section on Error! Reference source not found. for more details on what different smart contracts are employed on the platform.

The OConsent platform operates a Sidechain. Simply put, Sidechains are a separate blockchain with its own set of actors, e.g., validators and operators. The Sidechain frequently transfers assets to main chain and back. One of the key purposes, (and the same purpose of OConsent) is to capture the snapshot of the block headers to Main net to provide necessary guardrails against forking by bad actors on the sidechain.

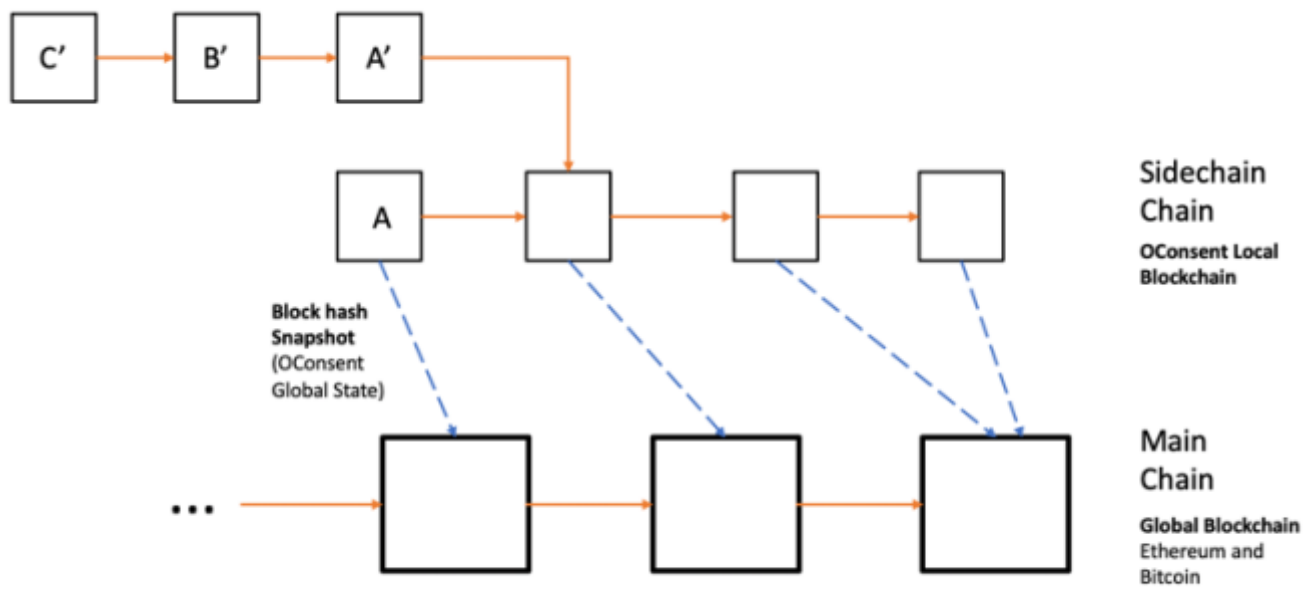

***Figure 12.*** *Sidechain interactions with a Main Chain and Security Implications.*

In the figure above, if malicious validators of the sidechain conspire and collude to produce a different and longer chain with C'-->B'-->A' after the Block A has been mined and the OConsent Local Block A has been snapshotted onto the Main Chain – the longer chain would be discarded by the sidechain participant.

Furthermore, where a OConsent Local Sidechain participant wants to download a consent proof from the Main Chain, he/she must lock the 'batch consents' on the main chain and provide a proof of the lock to the side chain. To unlock the 'batch consents' on the main chain, the participant must initiate an exit on the local sidechain and publish a proof of the exit after the 'batch consents' have been added to the local sidechain block.

## G. Local Blockchain Smart Contracts

Smart Contracts are key part of the OConsent Platform, ensuring autonomous and reliable executions of platform function. This section describes a non-exhaustive list of various Smart Contracts required for the platform.

1) *Managing Ownerships*

This smart contract is used to manage overall ownership of a Smart Contract. This smart contract is used to re-align ownership in case of Merged Identities of Data Controllers or when Data Subject accounts are in-active and assumed to be orphaned – requiring re-tagging to Platform accounts. This SC is used to limit certain functions to the owner of the contract only.

[4] This Hash Value is used as the OConsent Timestamp hash (trust anchor) and embedded as part of Agreement Proof.

**Outline of a sample NIST Randomness Beacon based Timestamp Proof**

See **Error! Reference source not found.** for the full body of the response. Only key sections of importance are highlighted below.

```
INPUT TIMESTAMP

Time of Beacon Pulse: 09/28/2020 23:25 +0800

------------------------------------------------------------------------------------

OUTPUT BEACON RECORD

URI: https://beacon.nist.gov/beacon/2.0/chain/1/pulse/1084642


Version: Version 2.0

Cipher Suite:0: SHA512 hashing and RSA signatures with PKCSv1.5 padding

Period:60000 milliseconds


Certificate Hash:

    MIIHYzCCBkugAwIBAgIQDnppkEkoPj8ZGd7VC7WwnjANBgkqhk [...] iG9w0BAQsFADBN

    qDdtoe3fwjVfFTUidnxZ1ISdqCAYOec=


Chain Index:1

Pulse Index:1084642

Time:2020-09-28T15:25:00.000Z


[...]


Signature:

    2D61EECD6E228ED7E81 [...] 623C098104CF51B7978724E829EC41AA51961D584

Output Value:

    CCDDD16135C36C673237328ECE38D01A3E1DAC817BB7005237088FA10502B6B1

    86291AD6059B09BC2B5B7744AA135BFDAB89FBE0E11E8FA1C99A665FB41CDF5B


Status:0: Normal
```

```
pragma solidity ˆ0.7.4;

///OConsentOwnership.sol

contract OConsentOwned {

    address public ownerCurr;

    event LogTransfer(address indexed

                ownerPrev, address indexed ownerNew);


    modifier ownerSole() {

    require (msg.sender == ownerCurr);

    _;

    }


    function OConsentOwned() public {

          ownerCurr = msg.sender;

    }


    function transferOwnership(address ownerNew) public

          ownerSole {

                require (ownerNew != address(0));

                LogTransfer(ownerCurr, ownerNew);

                ownerCurr = ownerNew;

    }

}
```

**Managing Ownerships**

```
pragma solidity ˆ0.7.4;

///OConsentOwnership.sol

contract OConsentOwned {

    address public ownerCurr;

    event LogTransfer(address indexed

                ownerPrev, address indexed ownerNew);


    modifier ownerSole() {

    require (msg.sender == ownerCurr);

    _;

    }


    function OConsentOwned() public {

          ownerCurr = msg.sender;

    }


    function transferOwnership(address ownerNew) public

          ownerSole {

                require (ownerNew != address(0));

                LogTransfer(ownerCurr, ownerNew);

                ownerCurr = ownerNew;

    }

}
```

*2) Time Based Consent Lease*

This contract is used to employ a time-bound lease of consent. E.g., As an end user I may limit my consent for marketing purposes for a period of 3 months only. In such a case, the Smart Contract will automatically cease to function after 3 months. This is one of the most powerful functionalities of the platform, that guarantees strict compliance to a Data Subject's consent.

This contract may also be expired for housekeeping and administrative purposes periodically. It may also be deprecated for security reasons as well.

```
pragma solidity ˆ0.7.4;

contract OConsentAutoDeprecate {

    uint expiresIn;



    function
OConsentAutomDeprecate(uint _days)
public {

            expiresIn = now + _days *
180 days;

    }

    function expiredAlready() internal
view returns (bool) {

            return now > expiresIn;

    }

    modifier upcomingDeprecation() {

            require(!expiredAlready());

            _;

    }

    modifier postDeprecation() {

            require(expiredAlready());

            _;

    }

    function grantConsent() public
payable upcomingDeprecation {

            // Code to execute

    }

    function withdrawConsent() public
view postDeprecation {

    // Code to execute

    }

}
```

*3) Linking Data Storage Repository*

This utility function is for separating the Data and the operational function; thereby enabling decoupling and easier upgrade of the operational function. The solidity code below is saved as *OConsentDataStorage.sol*. Other Smart Contract that executes the logic can refer to this contract file.

```
pragma solidity ˆ0.7.4.

///OConsentDataStorage.sol

contract OConsentDataStorage {

    mapping (bytes32 => uint)
uintDataSubjectStorage;

    function
getDataAddressUintValue(bytes32 key)
public constant

            returns (uint) {

            return
uintDataSubjectStorage[key];

    }

    function
setDataAddressUintValue(bytes32 key,
uint value) public {

            uintDataSubjectStorage[key]
= value;

    }

}
```

*4) Consent Agreement Updates Handler*

The following smart contract handles the changes in the Consent Agreement, by updating a new version of the smart contract. The SC tracks all the different version of the contracts and points to the latest version. This is an important SC which makes sure any agreement changes between the Data Subject and Data Controller are immediately reflected, including the revocation of access to Data Subject's private data.

```
pragma solidity ˆ0.7.4;

import "./OConsentOwnership.sol";

contract OConsentRegister is
OConsentOwned {

    address linkedContract;

    address[] previousLinks; /// list
of all previously linked contracts

    function OConsentRegister() public
{

            owner = msg.sender;

    }

    function changeLink(address
newLink) public

                    ownerSole() returns
(bool) {

            if(newLink !=
linkedContract) {


    previousLinks.push(linkedContract);

                    linkedContract =
newLink; /// add new contract as the
latest

                    return true;

            }

            return false;

    }

}
```

5) *Proxy Contract for 3rd Party Data Controllers*

This contract is used to provide a proxy for 3rd Party Data Controllers, who would inherit the same consent permissions as the Primary Data Controller. This proxy SC helps differentiate the different executions of the same proxy contract – thereby providing definitive audit trails.

```
pragma solidity ˆ0.7.4;

import "./OConsentOwnership.sol";



contract OConsentProxyRelay is
OConsentOwned {

    address public currVer;

    function OConsentProxyRelay(address
initialAddress) public {

            currVer = initialAddress;

            currOwner = msg.sender;

    }

    function updateContract(address
newProxyVer) public

    ownerSole() {

            currVer = newProxyVer;

    }

    /// Use this as a fallback

    function() public {


    require(currVer.delegatecall(msg.da
ta));

    }

}
```

H. *Fingerprinting on Global Blockchain*

Refer Figure 13.

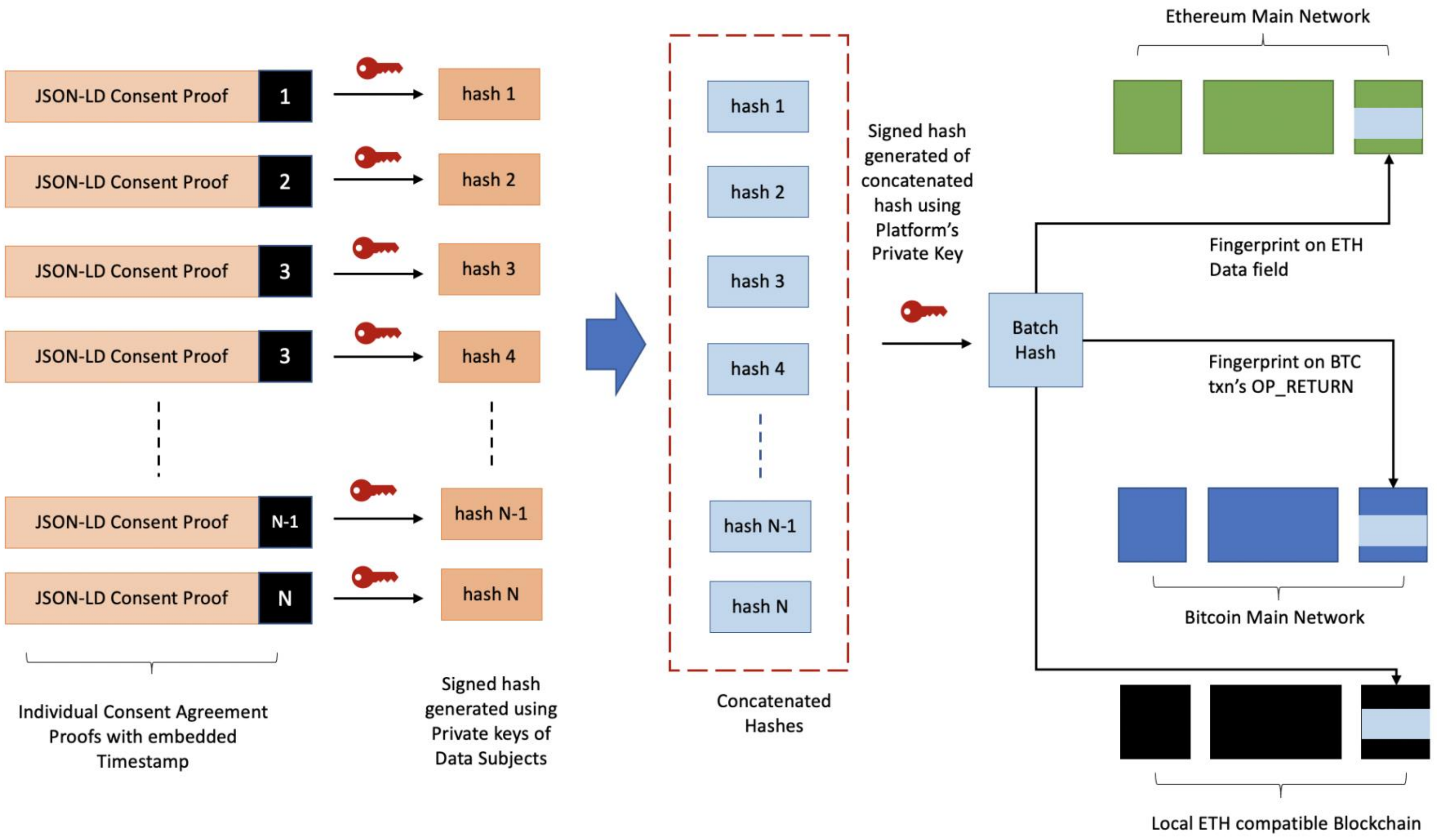

*Figure 13. How consent hashes are fingerprinted onto the public blockchains, Ethereum and Bitcoin*

1) *Fingerprinting on Bitcoin*
The Batch Hash is set in the OP_RETURN field of the Bitcoin Transaction. OConsent Platform can be configured to wait 'n' number of confirmations before the hash of the transaction is retrieved and embedded alongside the batch

2) *Fingerprinting on Ethereum*
The Batch Hash is embedded in the Extra Data field of the Ethereum Transaction.

3) *Fingerprinting on Local Blockchain*
As the local blockchain is an Ethereum compatible sidechain, the Batch Hash is embedded in the Extra Data field as well. Proofs are available at *https://oconsent.io/block/<BLOCK_NUM>*

4) JSON-LD Fingerprinting Proof
*Refer Figure 14.*

### I. *Anonymity on OConsent Platform*

OConsent Platform provides anonymity to Data Subjects (or End Users) in the following ways.

1) *Surrogate Identities*
The platform maintains a one-to-many mapping of Data Subjects primary and surrogate identities. This provides the DS's with the ability to anonymously share their data. A particular use case would be when a DS would not want to be tracked by advertisers using their real identities. It also provides the necessary guard against Data Controllers bypassing the OConsent Platform and the DS to collude among each other to share the data.

2) *Zero Knowledge Proofs*
There are multiple Zero Knowledge Proofs available, e.g., Zk-SNARKs, Zk-Starks, and Bulletproofs. Zk-SNARKs have been implemented successfully in production[lvii] and hence is the choice of Zero Knowledge Proof for the OConsent Platform. Zk-SNARKs [lviii] provides a proof construction whereby the Data Subject can prove possession of certain information without revealing that information, a without any explicit interaction between the prover (Data Subject) and verifier (Data Controller and Data Processor). Zk-SNARKs can help to verify proofs within a few milliseconds and "succinctly" provide proof within a few hundred bytes.

**Sample Use Case:** A possible use case of Zero Knowledge Proofs would be, if a Data Controller wants to know if a Data Subject is 18+ years old. Zk-SNARKs based non-interactive proofs may provide the answer without having to reveal the actual date of birth, thereby providing anonymity.

### J. *Interoperability Standard: Integration Request Formats*

OConsent platform uses the New Generation Access Control (NGAC) as a standard markup language for handling data access request:

- During the initiation phase of agreement proposal by DC
- For accessing additional data attributes access as an addendum to the consent agreement by the DC.
- For audit and proofs access by the Data Validators (DVs)
- Interoperations across other NGAC supported Data Storage Providers and Processors.
- Other data access requests.

NGAC is a reference implementation of the Policy Machine (PM) and has clear advantages over the XACML (Extensible Access Control Markup Language). NGAC computes decision through a linear algorithm over non-conflicting policies, thereby making it operationally efficient over XACML that requires collecting attributes and running computations (matching conditions, rules, and conflict resolutions) across a minimum of two different data stores - leading to extended complex computation steps. The proposed OConsent platform recognizes the clear advantages of NGAC over XACML and hence uses NGAC based markups to handle incoming consent and data access requests from Data Controllers. NGAC also includes a standardized set of administrative operations with a unified interface. It also provides the same interface for decision making function for accessing data assets, which is remarkably amiss in XACML.

Please refer to **Appendix B** for a Sample Java implementation of NGAC for OConsent.

### K. *Resolving Classification Conflicts*

OConsent incorporates a module for resolving conflicting rules or policies of a Data Subject's consent. Traditionally, Fuzzy Cognitive Maps have been used for modelling complex systems but are known to be marred by time lags between causes and observed effects. Consequently, Generalized Fuzzy Cognitive Maps (GFCM) and Generalized Rules Fuzzy Cognitive Maps (GRFCM) have been recently proposed to overcome such challenges.

For OConsent platform, Double Induction[lix] is proposed to be used. The idea behind Double Induction is that it induces unordered rules defined on the instances that are covered by the rules in conflict. By following this approach, new non-conflicting rules (because of separating the classes) are obtained by focusing on a smaller sub-space. This approach performs better over traditional Fuzzy Maps, Naïve Bayes, and frequency-based classifications. Double Induction method does include a higher computation cost but the same is offset by the remarkable accuracy it attains – which is one of the key proponents of having this module on the OConsent Platform.

```
{
      @context: https://w3id.org/oconsent/v1
      type: "OConsent – Fingerprinting Proof",

      signed_batch_hash_id: 11507a0e2f5e69d5dfa4...431b36fff21c437,

      fingerprints: [
          {
            bc_type: "BTC",
            block_id: 659800,
            URIs: [
                        https://www.blockchain.com/btc/block/0000000000000000000
                  ddb9e7d8747fa25e843b8f9bd13b18ba813349ce874a7
                  ]
           },

          {
            bc_type: "ETH",
            block_id: 011381576,
            URIs: [
                        https://etherscan.io/block/11381576,
                        https://etherchain.org/block/11381576
                  ]
           },

          {
            bc_type: "OCONSENT_ETH",
            block_id: 0000016,
            URIs: [
                        https://oconsent.io/block/0000016
                  ]
           },

      ]
}
```

*Figure 14. Sample JSON-LD OConsent Fingerprinting Proof*

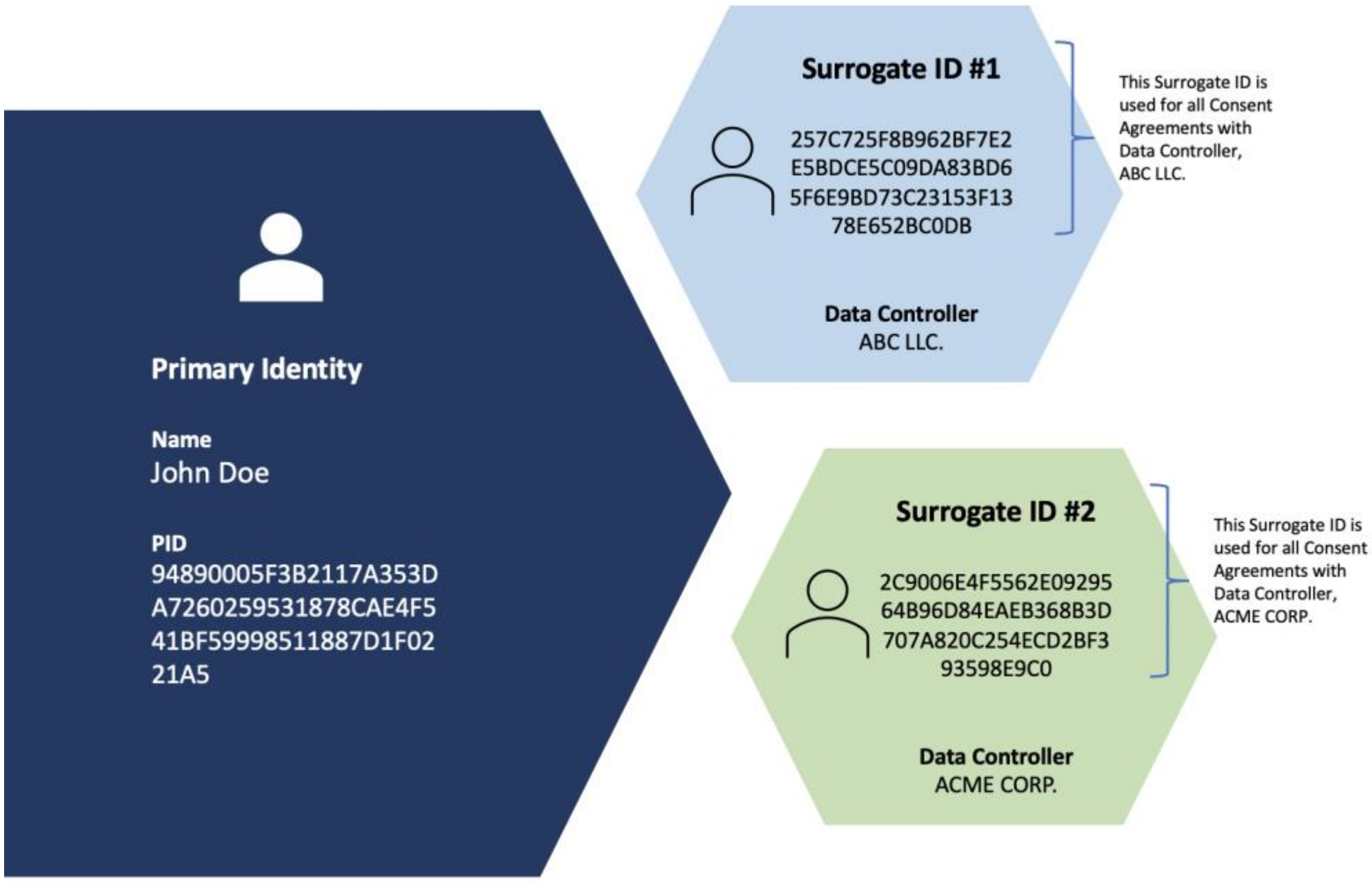

*Figure 15. One-to-Many map of Primary to Surrogate IDs to maintain anonymity*

## III. Summary

In this paper, I proposed the OConsent (Open Consent) framework that provides a comprehensive Consent Management System aligned to GDPR and other data privacy legislations using blockchain technology. The key goal of the platform is to provide a user-friendly solution that provides a one-stop solution for end users to manage their consent reliably and confidently. Optionally, the platform provides anonymity to the users using surrogate IDs or Zero Knowledge Proof – a first of its kind. Furthermore, a Double Induction based conflict resolution service is provided to better guide and advice Data Subjects (end users) while entering into Consent Agreements.

OConsent takes a practical approach to managing the Consent lifecycle with a Permissionless local sidechain. It provides multiple authoritative proofs for Consent receipt and validity for Auditors and Data Subjects. OConsent is also the only platform that implements a Trusted Timestamp proof to establish a non-repudiable point-in-time validity of a Signed Consent Agreement. OConsent also uses "multiple" Public Blockchains e.g., Bitcoin and Ethereum for fingerprinting the state of the local sidechain and thereby redundant proofs.

OConsent uses a standardized and most efficient access control policy mark-up language, NGAC. The platform has been designed to address scalability and performance needs from the initial. While the platform uses Blockchain technology, it balances the high throughput and low latency needs though an in-memory global state caching layer.

Additionally, costs can be controlled by offloading the same to Data Controllers and Data Processors. As the platform is based on a sidechain approach, and only global states are fingerprinted on Bitcoin and Ethereum – the operating costs would be considerably lower that solutions that imprint all consent agreements on the main chain.

## IV. Further Work

As a future work, I intend setup a working solution of the OConsent platform. The implementation would be open-sourced, contributions are welcomed - https://github.com/OConsent

Other alternatives to a sidechain implementation, e.g., Plasma and Hyperledger Besu – would be explored, as well as designing of an adaptive intelligent scheduler for fingerprinting on Bitcoin/Ethereum at the most optimal time of lower Ether costs.

Finally, I would look for monetization possibilities and payback to the Data Subjects for their consented data usage on the platform by Data Processors and Data Controllers.

### A. *Abbreviations and Acronyms*

| | |
|---|---|
| ABAC | Attribute Based Access Control |
| ADC | Auxiliary Data Controller |
| API | Application Programming Interface |
| AWS | Amazon Web Service |
| BC | Blockchain |
| CC | Chain Code |
| DAK | Data Access Key |
| DC | Data Controller |
| DLT | Distributed Ledger Technology |
| DS | Data Subject |
| DV | Data Validator |
| GCP | Google Cloud Platform |
| GCS | Google Cloud Storage |
| GDPR | General Data Protection Regulation |
| JSON | JavaScript Object Notation |
| JSON-LD | JSON-Linked Data |
| NGAC | Next Generation Access Control |
| NIST | National Institute of Standards and Technology |
| PM | Policy Machine |
| RBAC | Role Based Access Control |
| REST | Representational State Transfer |
| S3 | Simple Storage Service |
| SC | Smart Contracts |
| SDK | Software Development Kit |
| TSA | Time Stamping Authority |
| XACML | Extensible Access Control Markup Language |
| XML | Extensible Markup Language |
| Zk-SNARK | Zero Knowledge Succinct Non-Interactive Argument of Knowledge |

### B. *List of Figures*

# V. Appendices

## A. *Appendix A*

Sample NIST Randomness Beacon based Timestamp Proof

**INPUT TIMESTAMP**

Time of Beacon Pulse: 09/28/2020 23:25 +0800

-----------------------------------------------------------------------------------

**OUTPUT BEACON RECORD**

URI: https://beacon.nist.gov/beacon/2.0/chain/1/pulse/1084642

Version: Version 2.0

Cipher Suite:0: SHA512 hashing and RSA signatures with PKCSv1.5 padding

Period:60000 milliseconds

Certificate Hash:

MIIHYzCCBkugAwIBAgIQDnppkEkoPj8ZGd7VC7WwnjANBgkqhkiG9w0BAQsFADBN

MQswCQYDVQQGEwJVUzEVMBMGA1UEChMMRGlnaUNlcnQgSW5jMScwJQYDVQQDEx5E

aWdpQ2VydCBTSEEyIFNlY3VyZSBTZXJ2ZXIgQ0EwHhcNMjAwNjEwMDAwMDAwWhcN

MjEwNjExMTIwMDAwWjCBpTELMAkGA1UEBhMCVVMxETAPBgNVBAgTCE1hcnlsYW5k

MRUwEwYDVQQHEwxHYWl0aGVyc2J1cmcxNzA1BgNVBAoTLk5hdGlvbmFsIEluc3Rp

dHV0ZSBvZiBTdGFuZGFyZHMgYW5kIFRlY2hub2xvZ3kxEjAQBgNVBAsTCUlUTCAv

IENTRDEfMB0GA1UEAxMWZW5naW5lLmJlYWNvbi5uaXN0LmdvdjCCAiIwDQYJKoZI

hvcNAQEBBQADggIPADCCAgoCggIBAK3FLSzAVe4RmL17EfB9Ddir+bwP7PLhTqc+

ncUh3TRR4LMfRRKyt9TDYF8qm05MAKr738HP1Owsl5iQLNE8hsx4USk0WCfKb1LG

v6TajI6wFEFre4gTTYxmY9mjZ0ELe+xwtOaya5p9Cbq74CThcM11net5zAt4gd0d

N2uNjs9YPGuUi7OSc5zv7hL8ElzJwp76lemeGS2L7kDRuegMvEort7+5055a9mVN

sjhxy9vikutQBMUJlJbzwC6Zq5RLNuifFghysnIlrKw0CpACzxJ+3ZmpzhB2Swjp

J79h8oJMasQJw8Gsuw1u0rNrVV1rMrWUBYdn+5yTDd6XlzCfNGtxlaoExBEENt8T

lK0BRhI8vNv3H2GAy56WxYuhC4WD6t6b4KWuEUmV0myEdY98plPbPvLk/f3vUTJh

8qL2qop2NI+F28CXbByRg0i18i0uBRV7+oP2Hx5EzDk/GK9pEXSuQH6lIM5QMnb9

TRkHmE35dT7Zn1LxD9kGRgExWK7XQMHlJ030uV8L9PCwBimpwXAr/2G+6rzoBXq8

VIKIsbIw1AARCdx/JV8PYqzylat2cRZ3ap8TvaopFBFIcsAxhOyXRqKn5cyA/YL2

rR2BMt6XeQmnfhD8L5vJoqb/Q/NmnYnEX2/kX8FIVnjM+0hEFmvaH5Cz4qrJ/6oR
b8HXLryJAgMBAAGjggLkMIIC4DAfBgNVHSMEGDAWgBQPgGEcgjFh1S8o541GOLQs
4cbZ4jAdBgNVHQ4EFgQUvvas1JZiv0Jvtc1k0eI4eGlxvi0wIQYDVR0RBBowGIIW
ZW5naW5lLmJlYWNvbi5uaXN0LmdvdjAOBgNVHQ8BAf8EBAMCBaAwHQYDVR0lBBYw
FAYIKwYBBQUHAwEGCCsGAQUFBwMCMGsGA1UdHwRkMGIwL6AtoCuGKWh0dHA6Ly9j
cmwzLmRpZ2ljZXJ0LmNvbS9zc2NhLXNoYTItZzYuY3JsMC+gLaArhilodHRwOi8v
Y3JsNC5kaWdpY2VydC5jb20vc3NjYS1zaGEyLWc2LmNybDBMBgNVHSAERTBDMDcG
CWCGSAGG/WwBATAqMCgGCCsGAQUFBwIBFhxodHRwczovL3d3dy5kaWdpY2VydC5j
b20vQ1BTMAgGBmeBDAECAjB8BggrBgEFBQcBAQRwMG4wJAYIKwYBBQUHMAGGGGh0
dHA6Ly9vY3NwLmRpZ2ljZXJ0LmNvbTBGBggrBgEFBQcwAoY6aHR0cDovL2NhY2Vy
dHMuZGlnaWNlcnQuY29tL0RpZ2lDZXJ0U0hBMlNlY3VyZVNlcnZlckNBLmNydDAJ
BgNVHRMEAjAAMIIBBgYKKwYBBAHWeQIEAgSB9wSB9ADyAHcA9lyUL9F3MCIUVBgI
MJRWjuNNExkzv98MLyALzE7xZOMAAAFyn+jDnwAABAMASDBGAiEAlKqPJYbQEH9n
DfTW2dxjsXXb60RK2dXKYnjMFsWPLkwCIQCZWuTeEleSWUHYBNf3Q918GnBws6r2
6ZTsiqKCC+IY2gB3AFzcQ5L+5qtFRLFemtRW5hA3+9X6R9yhc5SyXub2xw7KAAAB
cp/ow8IAAAQDAEgwRgIhALlCMKPUYjj1kW8De07XcKdmux9YdLNAmNMOi6aOLlna
AiEA6B0xB2TRKzrebFXjVI2gFnzx5XMbEqAf1DQBCHeNXbcwDQYJKoZIhvcNAQEL
BQADggEBANGoFl6wzJzzoDmJ79QViGnCWFTtA2mkif5Z/sktAgKeAwsdER8WM0oi
1lDQmrkqmDV47OzbNKtgd2UOmTKHH6U+jvIwqg7jP1SpQN+HnfWNZzwjmKGJwLYA
e3RnzkufYtMSDYx3VVDrSiROT2AJ5W3i58+Wr9E35qAdx2nTGdyzGjbuzsr6qclw
7OBSignc00B6DOnQJT6PM1zbI24B7aCiPskXhlL2f+rBEFPPrBUGLlnRuwonCv4/
P4XdrILXwgECRrRO8v7U/KcyC0xYdEpMvw+gs+5f6RbGC7tXoe0+lI/irbVlJeyK
qDdtoe3fwjVfFTUidnxZ1ISdqCAYOec=

Chain Index:1

Pulse Index:1084642

Time:2020-09-28T15:25:00.000Z

Local Random Value:

A30D170D360FE8855BD7354D3FA7DB654FC104AE3A718433DE6155C0CAC1DFB1

```
    FC778D652D673D9FDC3586552E9647977F477AF3908ABA071C02B87ECD818246


External Source Id:

    0000000000000000000000000000000000000000000000000000000000000000

    0000000000000000000000000000000000000000000000000000000000000000


External Status Code:0


External Value:

    0000000000000000000000000000000000000000000000000000000000000000

    0000000000000000000000000000000000000000000000000000000000000000


Previous Output:

    5E02533A3855EF95F219A9F4017AB5B61AC9CF2289F540FCEA9505E5EA1D23D6

    498DC3ECC0C72E635211BE73673A79C42BEBAB41068EE97F0E2FC1538E17A07A


Hour:

    032D94B38419AB4071F8907A0C877707CA01C32705196B4A5173F909A266D2D0

    BCBD656E03CF9668E0F58F74B754A3B454A2E104388A10A689CAF73EE5506BB0


Day:

    2875F7EEA2B7AF715C8B0F077E18D40374C8CE8467775F8ED6BC7D19C4BC065E

    D51BE211E24111EA1C09F7124361DD39F57157C57550D6FE736C075E7EAE3E89


Month:

    55D5ABD00219290BF41190092C90E7DB429A80A468D4F6A643C1F09357FE820E

    C664A411D71A49E8680F78C5D962DC3EAF68A9F4031C29866E5D4468BB2C0F18


Year:

    CBC9AA97CDD5954218C585C89B061F356EF5F4158622C7CB38FBC317CA69C7AB
```

```
    E9E4379D4738B1076F7671C916C78AD0167A9ADB5A53E0CB20CC7F3D38736857

Precommitment Value:
    4E85EFADB2E0B74D53EC7062B9342C3477F1AFD8EBD7FEB58D16ADCDFEA67D37
    F0F862C4B27B79063EDA7869437EB910396057AFBF298777937E59DA3ADC0F5D

Signature:
    2D61EECD60911F8848835F6E0BFDE98B7BDB63C7C67BB4CF89F0FED32A30CD32
    CFD48F760EC36E2D5F88F23CA1499C32FC043199D05D7DCAAAF9719ABACDCA69
    22D5ECBC444E9B1F5526C08A3AE01C4373B9D940367058A126487E0EC6B1743F
    78DC71ACCD4BA7FEDEB9355ABB47D24A899E21A62A862B7FC6128062A6741F05
    1C7E7B284DA5B093E946C5A9901E58C5BB891539198FC5280AEF36626F7FF096
    1729E79D57CA72B20EFBC47DAB14C3EC770459FAFA06808181F958921CD86384
    0596EA7B43468DCF59C525264D063D65FBAA518184C48143A30DAFBAB2E10E51
    0D82F989F382E59CF6BF4467DF6E078D8EABA5803330C5322CD6E48F5C7B2377
    650B8A39055C4179573C88E03A0CBAA3D02754025DF73805BD08050AEB631268
    D70553800B58D7B32BD0DB344EE0293D129302993EE8862ACA7556EF845B2200
    3B30103FD53A830F3325308A581C20692EA6A3951BD440EC4912C90214F416C6
    DCBFFB96430B3B73C502E72D40CC94457252E1789ADA9732F3F628AB93F04F06
    CA0F9797D1273A43DFEE61C88105912724D7AFA275C9AFFFD65C620847EAF75C
    27BB20970A8D39F2B5D80A23EAE23B97557B24E1E01BD2EF658A551EAEA83BD2
    65BCE52198FC1DFEB92B8C622D44C5196D4FAA16B3413304855229941BC3C43F
    E228ED7E815808217C778E7623C098104CF51B7978724E829EC41AA51961D584

Output Value:
    CCDDD16135C36C673237328ECE38D01A3E1DAC817BB7005237088FA10502B6B1
    86291AD6059B09BC2B5B7744AA135BFDAB89FBE0E11E8FA1C99A665FB41CDF5B

Status:0: Normal
```

*B. Appendix B*

Sample Java Code for OConsent NGAC (New Generation Access Control) Implementation

```
package io.oconsent.ngac;

import gov.nist.csd.pm.decider.*;

import gov.nist.csd.pm.graph.*;

import gov.nist.csd.pm.prohibitions.model.Prohibition;

import java.util.*;

import static org.junit.jupiter.api.Assertions.assertTrue;



/*

The following snippet shows how an NGAC policy may be defined in Java, with Data
Subjects (DS),

Data Controllers (DC) and Data Processors (DP).

It also shows how multiple data assets may be linked to multiple Consent Agreements
and DC/DP.

*/



public class OConsentPolicy {

    protected OConsentPolicy() {

        // default constructor

    }

    public static class Builder {

        public static OConsentPolicy build() throws PMException {

            Random r = new Random();

            Graph g = new MemGraph();



            // Create Users

            Node oconsentUserM = g.createNode( r.nextLong(), "John Doe", NodeType.U,
null);



            // Create Admin User and attributes
```

```
            Node admin = g.createNode( r.nextLong(), "OConsent Admin", NodeType.UA, null);

            // Make user and admin

            g.assign(oconsentUserM.getID(), admin.getID());




            // Create objects

            Node dataAsset1 = g.createNode(r.nextLong(), "DataAsset1", NodeType.O, null);

            Node dataAsset2 = g.createNode(r.nextLong(), "DataAsset2", NodeType.O, null);

            Node dataAsset3 = g.createNode(r.nextLong(), "DataAsset2", NodeType.O, null);




            // Create the OConsentPolicy Class

            Node dataAssetOConsentPolicy = g.createNode( r.nextLong(), "DataAsset Access OConsentPolicy", NodeType.PC, null);




            // Create the data subject attribute

            Node dataSubjectX = g.createNode( r.nextLong(), “DataSubjects”, NodeType.OA, null);

            // Create Data Controller A and Data Processor B

            Node dataControllerA = g.createNode( r.nextLong(), "DataController", NodeType.OA, null);

            g.assign(dataControllerA.getID(), dataSubjectX.getID());




            Node dataProcessorB = g.createNode( r.nextLong(), "DataProcessor", NodeType.OA, null);

            g.assign(dataProcessorB.getID(), dataSubjectX.getID());
```

```java
            // Assing DataAssets to DataSubjects

            g.assign(dataAsset1.getID(), dataControllerA.getID());

            g.assign(dataAsset2.getID(), dataProcessorB.getID());

            g.assign(dataAsset3.getID(), dataProcessorB.getID());



            // Create agreements - these are tagged to Hashed Signed Agreements of the
OConsent Platform.

            Node consentAgreementGlobal = g.createNode( r.nextLong(),
"ConsentAgreements", NodeType.OA, null);



            Node agreementMarketing_1 = g.createNode( r.nextLong(),
"agreementMarketing", NodeType.OA, null);

            g.assign(agreementMarketing_1.getID(), consentAgreementGlobal.getID());

            Node agreementAnalytics_1 = g.createNode( r.nextLong(),
"agreementAnalytics", NodeType.OA, null);

            g.assign(agreementAnalytics_1.getID(), consentAgreementGlobal.getID());



            // Assign Data Assets to Consent Agreements

            g.assign(dataAsset1.getID(), agreementMarketing_1.getID());

            g.assign(dataAsset2.getID(), agreementMarketing_1.getID());

            g.assign(dataAsset3.getID(), agreementAnalytics_1.getID());



            // Assign the `Data Subject` and `Consent Agreements` objects attribute to
the `OConsentPolicy` policy class node.

            g.assign(dataSubjectX.getID(), dataAssetOConsentPolicy.getID());

            g.assign(consentAgreementGlobal.getID(), dataAssetOConsentPolicy.getID());



            //This will give user read and write on `Data Processor` and
`dataControllerA`

            g.associate(admin.getID(), dataControllerA.getID(), new
HashSet<>(Arrays.asList("r", "w")));

            g.associate(admin.getID(), dataProcessorB.getID(), new
HashSet<>(Arrays.asList("r", "w")));
```

```java
            //Create a NGAC Prohibition (for demonstration)

            Prohibition prohibition = new Prohibition();

            prohibition.addNode(oconsentUserM.getID(), agreementAnalytics_1.getID());



            Decider dec = new PReviewDecider(g);

            Set<String> permissions =
dec.listPermissions(oconsentUserM.getID(),dataAsset1.getID());

            assertTrue(permissions.contains("r"));

            assertTrue(permissions.contains("w"));



            return new OConsentPolicy();

        }

    }

}
```

Innovations in Theoretical Computer Science Conference on - ITCS '12. ACM. pp. 326–349. ISBN 9781450311151. S2CID 2576177

[xlvii] N. B. Truong, K. Sun, G. M. Lee and Y. Guo, "GDPR-Compliant Personal Data Management: A Blockchain-Based Solution," in IEEE Transactions on Information Forensics and Security, vol. 15, pp. 1746-1761, 2020, doi: 10.1109/TIFS.2019.2948287.

[xlviii] JSON-LD, "JSON for Linking Data", Retrieved, Nov 2, 2020, from https://json-ld.org/

[xlix] General Data Protection Regulation (GDPR), "Art. 4 GDPR Definitions", Retrieved Sep 10, 2020, from https://gdpr.eu/article-4-definitions/, Article 4 point (1)

[l] General Data Protection Regulation (GDPR), "Art. 4 GDPR Definitions", Retrieved Sep 10, 2020, from https://gdpr.eu/article-4-definitions/, Article 4 point (7)

[li] ITL Bulletin for November 2016, "Exploring the next generation of access control methodologies", NIST, Retrieved Nov 5, 2020, from https://tsapps.nist.gov/publication/get_pdf.cfm?pub_id=922332

[lii] Computer Security Resource Center, "Policy Machine", NIST, Retrieved on Nov 5, 2020, from https://csrc.nist.gov/Projects/Policy-Machine

[liii] Andreas M. Antonopoulos. 2014. Mastering Bitcoin: Unlocking Digital Crypto-Currencies (1st. ed.). O'Reilly Media, Inc., Chapter 6 The Bitcoin Network

[liv] Bitcoin Explorer, "Block 659792", Blockchain.com, Retrieved on Dec 4, 2020, from https://www.blockchain.com/btc/block/0000000000000000000aa23344fcefaafa535d40f3f6185aa71c05f361a50067

[lv] Drand, "About Drand", Retrieved Nov 25, 2020, from https://drand.love/about/

[lvi] Cloudflare Inc., "Cloudflare Randomness Beacon docs", Retrieved Nov 25, 2020 from https://developers.cloudflare.com/randomness-beacon/

[lvii] ZCash, "What are zk-SNARKs?", Retrieved Nov. 20, 2020, from https://z.cash/technology/zksnarks/

[lviii] Reitweissner C., "ZKSNARKS in a nutshell", Ethereum Blog. Retrieved on Nov. 20, 2020, from https://blog.ethereum.org/2016/12/05/zksnarks-in-a-nutshell/

[lix] Tony Lindgren and Henrik Boström. 2004. Resolving rule conflicts with double induction. Intell. Data Anal. 8, 5 (October 2004), 457–468.